\documentstyle[12pt]{JHEP3}
\newcommand{\be}[1]{ \begin{equation}\label{#1} }
\newcommand{\ee}{\end{equation}}
\newcommand{\bea}[1]{\begin{eqnarray}\label{#1} }
\newcommand{\eea}{\end{eqnarray}}
\newcommand{\eq}[1]{(\ref{#1})}

\def\ZZZ{{\hskip-3pt\hbox{ Z\kern-1.6mm Z}}}
\def\zzz{{\hskip-3pt\hbox{ z\kern-1mm z}}}
\newcommand{\Phit}{\widetilde{\Phi}}

\title{On the dyon partition function in ${\cal N}=2$ theories}
\author{Justin R. David, \\
 Harish-Chandra Research Institute, \\
Chhatnag Road., Jhunsi, \\
Allahabad 211019, India.
\\
\email{justin@hri.res.in}
}

\abstract{
We study the entropy function of two ${\cal N}=2$ string compactifications
obtained as freely acting orbifolds of ${\cal N}=4$ theories : the STU model
and the FHSV model. The Gauss-Bonnet term
for these compactifications is known precisely. 
We apply the entropy function formalism including the 
contribution of this four derivative term
and evaluate the entropy of dyons to the first subleading order in charges 
for these models. 
We then propose a partition function
involving the product of three Siegel modular forms of weight 
zero which reproduces the degeneracy of  dyonic black holes in the STU model
to the first subleading order in charges. The proposal
is invariant under all the duality symmetries of the 
STU model.
For the FHSV model  we write down an approximate partition function
involving  
a Siegel modular form of weight four 
which captures the entropy of dyons in the 
FHSV model in the limit when electric charges are much larger than magnetic
charges.
}

\begin{document}
\baselineskip 3.5ex

\section{Introduction}

Recent studies  has led to a good understanding of
the spectrum of $1/4$ BPS dyonic 
states in a class of ${\cal N}=4$ supersymmetric string
theories \cite{Dijkgraaf:1996it,LopesCardoso:2004xf,Shih:2005uc,Gaiotto:2005hc,
Shih:2005he,Jatkar:2005bh,David:2006ji,Dabholkar:2006xa,
David:2006yn,David:2006ru,David:2006ud,Dabholkar:2006bj,Banerjee:2007ub}.
These theories are a class of generic ${\cal N}=4$ supersymmetric
$\ZZZ_N$ orbifold of type IIA string theory on $K3\times T^2$ or
$T^6$. In each example studied so far, the statistical entropy computed
by taking the logarithm of the degeneracy of states agrees with the
entropy of the corresponding black hole for large charges.
This is not only in the leading order, but also in the first
sub-leading order \cite{LopesCardoso:2004xf,Jatkar:2005bh,
David:2006yn,David:2006ru,David:2006ud}.
On the black hole side this requires taking into account the effect of the
Gauss-Bonnet term in the low energy effective action of the theory,
and the use of the Wald's generalized formula for the black hole entropy in the
presence of higher derivative corrections 
\cite{Wald:1993nt,Iyer:1994ys,Sen:2005wa}.
These $1/4$ BPS dyons are known to have regions of marginal stability,
the degeneracies of the dyons jump across these regions of marginal stability.
These changes are precisely captured by
the same dyon partition function but with different 
choices of the integration contour, the moduli dependence of the
contour of integration is known precisely 
\cite{Sen:2007vb,Dabholkar:2007vk,Sen:2007pg,Cheng:2007ch,
Mukherjee:2007nc,Sen:2007nz,Mukherjee:2007af}. 
For a recent review on these topics see \cite{Sen:2007qy}.

So far similar studies in ${\cal N}=2$ theories has been lacking. 
Because of reduced  supersymmetry the Gauss-Bonnet term
in the low energy effective action is not just a term dependent on the 
axion and dilaton, but also on the other moduli of the vector multiplets.
The precise agreement 
of the asymptotic degeneracy of the dyons in ${\cal N}=4$ 
theory to the first sub-leading order depended crucially on just
the axion-dilaton dependence of the Gauss-Bonnet term. 
Any similar proposal for dyons in ${\cal N}=2$  theories should
address the question the moduli dependence of the other vector multiplets.

The study of 1/2 BPS dyons in a 
generic ${\cal N}=2$ theory would be a hard task. In this paper we focus
on two ${\cal N}=2$ theories which are closely related to 
${\cal N}=4$ theories. 
The first  theory is 
the STU model obtained by a  freely acting orbifold action of
type IIA on $T^4\times T^2$. There are three vector multiplets in this
theory, the $S$, $T$ and the $U$.
This model was constructed and studied in \cite{Sen:1995ff,Gregori:1999ns}, 
the coefficient of the Gauss-Bonnet term in this model is known exactly 
and it is a sum of contributions from the S, T and the U-moduli.
In this case the exact moduli space of the vector multiplets is 
given by
\be{vecstum}
{\cal M}_V = \left. \frac{SU(1,1)}{U(1)}\right|_S \times 
\left. \frac{SU(1,1)}{U(1)}\right|_T \times 
\left. \frac{SU(1,1)}{U(1)}\right|_U.
\ee
where each $SU(1,1)/U(1)$ factor parameterizes the respective moduli.
We  study  the entropy function for this model and evaluate the expectation
values of the moduli at the attractor point. 
This enables us to extract the S, T, U duality invariants constructed 
out of bilinears of the electric and magnetic charges
which characterize the entropy at the next leading order.
Using these invariants we  propose a partition function for dyons
in this model. This partition function involves the product of three
Siegel modular forms of weight zero, it has all the duality symmetries of the
STU model. The statistical entropy obtained from the partition function 
reproduces correctly the entropy of a dyonic black hole to the 
first sub-leading order for large values of charges \footnote{
For the  reader who is interested in the formula for the partition function,
please  see \eq{stupart}.}.

Another well known example of a ${\cal N}=2$ model closely related 
to a parent ${\cal N}=4$ theory is that of the self mirror
Calabi-Yau constructed in \cite{Ferrara:1995yx}.
On the type IIA side it is constructed by a freely acting orbifold of 
type IIA on $K3\times T^2$, the resulting Calabi-Yau
is  a self-mirror manifold. Therefore the moduli space of this theory is 
known exactly. The vector multiplet moduli space is given by 
\be{vecmoduli}
{\cal M}_V = \frac{SU(1,1)}{U(1)}  \times \frac{SO(2,10)}{S0(2) \times SO(10)}.
\ee
One of the important property of this moduli space is that the dilaton-axion
moduli factorizes from the rest of the vector multiplet moduli.
The dilaton-axion parameterizes the coset $SU(1,1)/U(1)$ while the 
rest of the moduli(the T-moduli)
 parameterize the coset $SO(2,10)/SO(2)\times SO(10)$. Because of this 
factorization, the coefficient of the Gauss-Bonnet term can be computed 
exactly \cite{Harvey:1996ts}. From the analysis of the entropy function for 
the FHSV model and the attractor values of the T-moduli 
we  show that this factorization allows one to 
parametrically separate the subleading
contribution to the entropy of dyons in this model into two parts. 
The contribution from the axion-dilaton dependence of the Gauss-Bonnet 
term is dominant when the electric charges are much larger than the 
magnetic charges. We then write down a Siegel modular form of weight 
4 which captures the dependence of the entropy function on the axion-dilaton
moduli.

The organization of the paper is as follows: In section 2. we review the 
construction of the STU model
as well as the FHSV model and recall the 
the coefficient of the Gauss-Bonnet term in these models. 
Section 3. contains the entropy function 
analysis of these models. We explictly solve for the attractor moduli 
in both these models. This enables us to evaluate the 
sub-leading contribution
to the black hole entropy of dyons from the coefficient of the Gauss-Bonnet
term in these models. It also helps us to determine the charge bilinears
characterizing the entropy at the subleading order. 
In section 4. we propose a partition function for dyons in the STU model
and show that it has the  required duality invariance and reproduces 
correctly the entropy of a dyonic black hole to the first sub-leading 
order for large values of the charges.
In section 5. we write down an approximate partition function for
dyon in the FHSV model which captures the entropy of the corresponding 
black hole for large values of the charges but with electric charges
much larger than the magnetic charges.
The appendices contain  
conventions regarding the 't Hooft symbols for SO(2,2) and 
the properties of the Siegel modular forms used in the paper.
Appendix B shows the systematic method by which 
the attractor equations for the FHSV model is solved. This procedure 
can be used for any model which has the following vector multiplet
moduli space
\be{genvecmodi}
{\cal M}_V = \frac{SU(1,1)}{U(1)} \times \frac{SO(2,n)}{SO(2) \times SO(n)}.
\ee 

\section{Two ${\cal N}=2$ theories}

In this section we review the construction of the two ${\cal N}=2$ models
that 
we will be studying.  The dyons which will be the focus of our interest 
preserve $1/2$  of the $8$ supersymmetries of these theories.

\subsection{The STU model}

This model is best described in terms of a freely acting $\ZZZ_2\times
\ZZZ_2$  orbifold of 
type IIB \cite{Sen:1995ff,Gregori:1999ns} (example C in \cite{Sen:1995ff}).
Consider Type IIB compactified on $T^4\times S^1 \times \tilde S^1$, 
the first $\ZZZ_2$, $g_1:$ acts
by $(-1)^{F_L}$ together with a half shift on $S^1$, 
the second $\ZZZ_2$, $g_2$ acts
as an inversion of the coordinates on $T^4$ together with a half shift 
on $\tilde S^1$. The theory with only with the  
$g_1$ action is the same as the 
one considered in \cite{David:2006ru}, this is a ${\cal N}=4$ theory. 
The second $\ZZZ_2$ action $g_2$ further breaks the supersymmetry down
to ${\cal N}=2$. 
This theory has $3$ vector multiplets and $4$ hyper multiplets.
The $T$-duality symmetry of this theory is 
$SO(2,2; \ZZZ)$. In the type IIA description of this orbifold the 
dilaton belongs to the 
hypermultiplet, while in the type IIB description \footnote{
Here the duality symmetry 
which relates the type IIA and the type IIB description of the theory
is not the conventional T-duality but the one
which is part of the U-duality group and
is a strong-weak duality symmetry \cite{Sen:1995ff}}
it belongs
to the vector multiplet \cite{Sen:1995ff,Gregori:1999ns}. 
Therefore the vector multiplet moduli space is not corrected
by quantum corrections and it is given by
\be{vecmstu1}
{\cal M}_V = \left. \frac{SU(1,1)}{U(1)} \right|_{S} \times
\left. \frac{SU(1,1)}{U(1)} \right|_{T}\times \left. \frac{SU(1,1)}{U(1)}
\right|_{U}, 
\ee
where $S$ refers to the axion dilaton moduli and $T$ and $U$ refer to the 
K\"{a}hler and complex structure of the torus $S\times \tilde S$. This theory
is invariant under the symmetry $\Gamma_S(2)\times \Gamma_T(2)\times
\Gamma_U(2)$ where $\Gamma(2)$ is defined as
\be{defgamm2}
\left(\begin{array}{cc}
a & b \\
c & d 
\end{array} \right)
\quad ad - bc =1, \quad b, c \in 2\ZZZ,\; a, d \in 2\ZZZ+1.
\ee
It also has the triality invariance
$S\leftrightarrow T\leftrightarrow -1/U$.

The dyons we consider are the twisted sector 
dyons of the parent ${\cal N}=4$ theory which
are preserved by the second orbifold projection $g_2$. Thus these dyons have 
electric, and magnetic charges  only along $S^1$ and $\tilde S^1$ directions.
Since these dyons are $1/4$ BPS states in the parent  theory they 
will preserve $1/2$ of the supersymmetries of the daughter theory.  
For the purposes of obtaining the subleading corrections to the entropy of the 
dyons we will need the 
coefficient of the Gauss-Bonnet term. The Gauss-Bonnet term is given
 by following combinations of 
4 derivative terms made of the curvature tensor
\be{defgb}
R_{\mu\nu\rho\sigma}R^{\mu\nu\rho\sigma} - 4 R_{\mu\nu} R^{\mu\nu} + R^2.
\ee
The coefficient of this term for the STU model  was evaluated in 
\cite{Gregori:1999ns},
this is given by
\bea{cofstugb}
 & & \frac{1}{128\pi^2} \left(  4\ln\left( \frac{M_p^2}{p^2} \right) -
4\log \left[\vartheta_2 (\tau)^2\vartheta_2(-\bar\tau)^2 (\tau_2) \right] 
\right. \\ \nonumber
& & \left. - 4\log \left[\vartheta_2(y^+)^2  \vartheta_2(-\bar y^+)^2
    (y^+_2)
 \right]
-4\log \left[\vartheta_4(y^-)^2\vartheta_4(-\bar y^-)^2 (y^-_2) \right]
\right),
\eea
where $\tau$ refers to the complex combination of the  axion 
and dilaton moduli given by
\be{defs}
\tau = -a + i S,
\ee
where $a$ is the axion and $S= \exp(-2\phi)$ is the dilaton.
 $y^+$ and 
$y^-$ refer to the K\"{a}hler and complex structure of the
$S\times\tilde S$ torus, which we have denoted as the $T$ and $U$ 
moduli
\footnote{Throughout the paper, the subscripts $1$ and $2$ on 
complex moduli refer to its real and imaginary parts respectively.}.
The normalization of the  Gauss-Bonnet term is determined as follows 
\cite{Sen:2005pu}: Let
the axion-dilaton dependence of the Gauss-Bonnet term be given by
\be{normgb}
-\frac{{\cal K}}{128\pi^2} 
\ln\left[ \tau_2 f(\tau) f(-\bar\tau) \right].
\ee
The functional dependence of the term in the square  bracket is invariant 
under the S-duality symmetry  $\Gamma(2)_S$.
Then the coefficient 
${\cal K}$ is given by the number of harmonic $p$ forms of $T^4$ left
invariant under the action of $g_1$ and $g_2$
weighted by $(-1)^p$. $g_2$ projects retains
only the even forms which results in $8$ forms. Since $g_1$ reduces the 
supersymmetry by projecting out the left moving fermions, it project out
the $3$ self-dual 2-forms and a combination of the zero form and $4$ form.
Thus 
out of the $8$ forms $4$ are left invariant. We therefore 
conclude  ${\cal K}=4$. Once
this normalization is determined the dependence of the axion-dilaton 
moduli and the other moduli is obtained from the calculation of 
\cite{Gregori:1999ns}, equation (5.3). The dependence of the 
coefficient of the Gauss-Bonnet term on $\ln(M_p^2/p^2)$ 
where $M_p$ is the Planck's constant and $p^2$ is the 
graviton momentum is due to the 
trace anomaly of the theory.
Though the coefficient of the trace anomaly does 
not play a role in the 
evaluation of the sub-leading contribution to the entropy, its 
contribution to the coefficient of the Gauss-Bonnet term  
for a general ${\cal N}=2$ theory is given by
\be{coeftracean}
-\frac{1}{128\pi^2} \left( \frac{23 + n_h - n_v}{ 6}\right) \ln 
\left( \frac{M_p^2}{\Lambda^2}\right),
\ee
where $n_h$ is the number of hypermultiplets and $n_v$ is the number
of vector multiplets, for the STU model $n_h-n_v=1$. 
Note that the coefficient of the Gauss-Bonnet term \eq{cofstugb} is
invariant under the triality symmetry 
$\tau \leftrightarrow y^+\leftrightarrow -1/y^-$ as well as
$\Gamma_S(2) \times \Gamma_T(2) \times \Gamma_U(2)$, which is 
the symmetry of the model. 
This completes the explantation of all the terms in \eq{cofstugb}.

\subsection{The FHSV model}

For our purposes it is easiest to describe the FHSV model as a freely 
acting orbifold of the heterotic $E_8\times E_8$ theory.
Consider the heterotic string on the following 
even, self-dual Lorentzian lattice $\Gamma^{(22,6)}$ of the form
\be{hetlat}
\Gamma^{(9,1)} \oplus \Gamma^{(9,1)}\oplus\Gamma^{(1,1)}\oplus
\Gamma^{(1,1)}\oplus\Gamma^{(2,2)}.
\ee
Here the two $\Gamma^{(9,1)}$ factors are isomorphic. 
We now orbifold by a $\ZZZ_2$ action which exchanges the first
two factors and act as $-1$ on the third $\Gamma^{(1,1)}$ and 
the last $\Gamma^{(2,2)}$ together with the half shift in 
the fourth $\Gamma^{(1,1)}$ factor. 
The T-duality group of this heterotic string is $SO(2,10 ;\ZZZ)$,
the moduli space of this theory is known exactly \cite{Ferrara:1995yx}. 
For purposes of this paper it is sufficient to focus on the vector multiplet
moduli space. There are $11$ vector multiplets in this
theory and it moduli space  is given by
\be{vmodfshv}
{\cal M}_V = \frac{SU(1,1)}{U(1)} \times \frac{SO(2, 10)}{SO(2)\times SO(10)}.
\ee
Note that the coset $SU(1,1)/U(1)$ which is parameterized by the 
axion-dilaton moduli
factorizes from the rest of the vector multiplet 
moduli, we call these moduli the T-moduli. The $S$ duality of this 
theory is $\Gamma_S(2)$.  

We focus on a class of 
dyon configurations which preserves $1/2$ of the remaining symmetry
in the FHSV model. These class of dyons are
 obtained by embedding the $1/4$ BPS dyon of
the parent ${\cal N }=4$ theory  such that the dyon configuration
is preserved by the 
FHSV orbifold action. 
We now describe the charge configuration of such a dyon:
Let us call the unprojected combination of the two $\Gamma^{(9,1)}$ lattices as
$\Gamma^{(9,1)}_I$, we denote the fourth 
$\Gamma^{(1,1)}$ as $\Gamma^{(1,1)}_S$.
We focus on dyons which have electric and magnetic charges values on the
lattices $\Gamma^{(9,1)}_I\oplus \Gamma^{(1,1)}_S$ which have twisted sector 
charges along the $\Gamma^{(1,1)}_S$ lattice. 
It is clear that such charge configurations are preserved by the 
FHSV orbifold action,
since these configurations do not have charges in the 
lattices $\Gamma^{(1,1)}\oplus \Gamma^{(2,2)}$ on which the 
FHSV orbifold acts as an inversion.

We will  study the macroscopic entropy of these dyons using 
the entropy function formalism to the first subleading term. 
For this purpose we will need the coefficient of the Gauss-Bonnet term
in this model. The dilaton-axion and the T-moduli 
dependence of the Gauss-Bonnet coefficient of this
model was evaluated exactly in  \cite{Harvey:1996ts}. 
We now review their result:
The coefficient of this term for ${\cal N}=2$ compactifications is given 
by 
\be{cofgb}
\frac{1}{128\pi^2} \left( 4   \ln(\frac {M_p^2}{p^2} ) 
 +  {\cal F}_1 (\tau,  \bar\tau, y, \bar y )\right).
\ee  
In \eq{cofgb} 
${\cal F}_1(\tau , \bar\tau)$ is the modular function determined by the 
threshold calculation and $\tau$ is the complex combination of the 
axion and the dilaton given in \eq{defs}. 
For the FHSV model, the $\tau$  and the T-moduli dependence of the 
threshold corrections is given by \cite{Harvey:1996ts}
\bea{threshfs}
{\cal F}_1(\tau, \bar \tau, y, \bar y) &=& 
-12\log \left[ \eta(2\tau)) (\eta(-2\bar\tau)  (\tau_2) \right]\cr
& &  - \log \left[ \Phi_{{\rm BE}} (y)  
\Phi_{{\rm BE}} (-\bar y)  ( 4 y^{+}_2 y^{-}_2 - \vec y_2^2)^4 \right],
\eea
where $\Phi_{{\rm BE}}(y)$ is the Borcherds-Enriques 
modular form of weight four on the 
discrete group $E_8\times \Gamma^{1,1}(-2)\times \Gamma^{1,1}(-2)$.
$y$ refer to the $10$ complex T-moduli, 
$y=\{ y^+, y^-, y^i\}, i = 1, \cdots 8$.  
The subscript $2$ on the various moduli in 
\eq{threshfs} stands for the imaginary values
the  moduli. For the purposes of this paper will not need the details of the 
Borcherds-Enriques form, please see \cite{Harvey:1996ts} for the
details. 
Just as in the STU model we fix the normalization of the 
coefficient of the Gauss-Bonnet term by examining the 
axion-dilaton dependence. This is given by
\eq{normgb}, here ${\cal K}$ is the number of p-forms invariant on 
the dual description of the the FHSV orbifold in type IIA theory on 
$K3$ weighted by $(-1)^p$. This turns out to be $12$, the FHSV orbifold
action projects out $12$ of the $24$ forms of $K3$. 
The dependence of the axion 
dilaton and the remaining T-moduli dependence is then read out from
\cite{Harvey:1996ts}. The trace anomaly dependence is given 
by \eq{coeftracean} with $n_v =11$ and $n_h=12$ for the FHSV model.

\section{Entropy function for dyons}

In this section we will evaluate the first sub-leading contribution to the 
Hawking-Bekenstein entropy due to Gauss-Bonnet term in the 
effective action for dyons in both the STU model and the FHSV model
 discussed in the previous
section. We will follow the entropy function approach developed in 
\cite{Sen:2005iz}. In this approach 
to find the first sub-leading contribution to the entropy we will need both the
coefficient of the Gauss-Bonnet term as well as the attractor values of  
vector multiplet moduli at the two derivative level.  
We first evaluate the attractor values of all the vector multiplet moduli
to both these models and then substitute these values in the 
coefficient of the respective Gauss-Bonnet term in these models to 
evaluate the sub-leading contribution to the Hawking-Bekenstein entropy.

To make the discussion self contained we briefly review the 
entropy function formalism. 
Consider an extremal black hole solution in any of the ${\cal N}=2$
theories of interest in this paper. The near horizon geometry of this
black holes is given by 
\bea{defsoln}
ds^2 = v_1 \left( -r^2 dt^2 + \frac{dr^2}{r^2} \right) + v_2 (d\theta^2 + 
\sin^2 \theta d\phi^2 ), \cr
S =u_S, \qquad a=u_a, \qquad \tilde M_{ij} = u_{M ij}, \cr
F_{rt}^{(i)} = e_i,  \quad 
F_{\theta\phi}^{(i)} = \frac{p_i}{4\pi}.
\eea
Here $i =1, \cdots n+2 $ and the near horizon vector multiplet moduli,
$n=10$ for the FHSV model and  $n=2$ for the STU model.
$u_{M {ij}}$ satisfies 
\be{modcond}
u_M^T L u_M = L, \quad u_M^T=u_M,
\ee
where $L$ is the Lorentzian metric with $n$ negative signatures and 
$2$ positive signatures; $L = {\rm Dia} ( -1, -1, \cdots -1, +1, +1 )$. 
Substituting the solution \eq{defsoln} in the supergravity action we obtain
\cite{Sen:2005iz}
\bea{nhsugra}
& &f(u_s, u_a, u_M, \vec v, \vec e, \vec p) \equiv
\int d\theta d\phi \sqrt{-{\rm det} G } {\cal L} \\ \nonumber
&=& \frac{1}{8} v_1v_2 u_S \left[
-\frac{2}{v_1} + \frac{2}{v_2} + \frac{2}{v_1^2} e_i ( L u_M L)_{ij} e_j
- \frac{1}{8\pi^2 v_2^2 } p_i (Lu_M L)_{ij} p_j +
\frac{u_a}{\pi u_S v_1 v_2} e_i L_{ij} p_j 
\right].
\eea
We then obtain the electric charges $q_i$ given by
\be{elchrg}
q_i \equiv \frac{\partial f}{\partial e_i} 
= \frac{v_2 u_S}{2 v_1} ( L u_M L)_{ij} e_j + \frac{u_a}{8\pi} L_{ij} p_j.
\ee
Evaluating the Legendre transform with respect to the variables $e_i$ we 
obtain the entropy function $F$
\bea{entfn}
F(u_S, u_a, u_M, \vec v, \vec q, \vec p) 
&\equiv& 2\pi ( e_i p_i - f(u_S, u_a, u_M, \vec v, \vec e, \vec p) ) \cr
&=& 2\pi \left[ \frac{u_S}{4} ( v_2 -v_1) + \frac{v_1}{v_2 u_S} 
q^T u_M q + \frac{v_1}{64\pi^2 v_2 u_s} ( u_S^2 + u_a^2) L u_M L p \right. \cr
&\;& \left. - \frac{ v_1}{4\pi v_2 u_S} u_a q^T u_M L p \right].
\eea
We define the charge vectors
\be{defnechr}
Q_i = 2 q_i, \qquad P_i = \frac{1}{4\pi} L_{ij} p_j,
\ee
so that $P_i$ and $Q_i$ are integers. Substituting these definitions of 
charges in the entropy function we obtain
\be{neentfn}
F = \frac{\pi}{2} \left[ u_S( v_2 -v_1) + \frac{ v_1}{v_2 u_S} 
\left( Q^T u_M Q + ( u_S^2 + u_a^2) P^T u_M P - 2 u_a Q^T u_M P \right)
\right].
\ee
Now to evaluate the Hawking-Bekenstein entropy we
 need to find the the extremum of the above function with respect to 
the moduli $u_S, u_a, u_{M ij}, v_1, v_2$. 
For this we can first minimize the entropy function in \eq{neentfn} with 
respect to $v_1$ and $v_2$. It is easy to see from the dependence of the
entropy function on $v_1$ and $v_2$, the minimum occurs at $v_1 = v_2$. 
We then organize the axion and the dilaton moduli in terms of the complex
scalar $\tau$ defined as 
\be{deftau}
\tau = -u_a +  i u_S.
\ee
Using $\tau$ and the parameterization and $v_1=v_2$, the entropy function
\eq{neentfn} reduces to 
\be{entfns}
F = \frac{\pi}{2\tau_2} \left[ ( Q + \tau P)^{T} u_M (Q +\tau P) \right].
\ee
In the next two subsections we use the above entropy function and 
to further minimize with respect to the T-moduli parameterized by
$u_M$ and the axion dilaton moduli $\tau$ for the STU model and the
FHSV model. In both the cases we obtain the attractor values of these
moduli in terms of the charges explictly. We then use the attractor 
values of the moduli to evaluate the subleading correction due to the
Gauss-Bonnet term.

\subsection{The two derivative entropy function: STU model}

For the case of the STU model the moduli matrix $u_M$ is a $4\times 4$ matrix
which parameterizes the coset $SO(2,2)/(SO(2)\times SO(2))$. It satisfies
the conditions
\be{condstu}
u_M^T = u_M, \qquad u_M^TL u_M = u_M,
\ee
where $L$ is the diagonal metric $L = {\rm Dia} ( 1, 1, -1, -1)$.
From the conditions in \eq{condstu} 
it can be easily seen that there are $4$ independent variables 
which parameterize the matrix $u_M$. A convenient parameterization is 
as follows:
We first introduce 4 complex numbers satisfying
\be{consteq1}
w_1^2 + w_2^2 - w_3^2 - w_4^2  = 0, 
\ee
together with the identification $w_I \sim c w_I$ where $c$ is a complex
number. Thus $w_I$'s parameterize the coset $SO(2,2)/(SO(2) \times SO(2))$. 
Using the scaling symmetry we can solve the constraint \eq{consteq1} by
introducing complex numbers $y^+$ and $y^-$ and 
writing $w_I$'s as 
\bea{relwystu}
& w_1 = -1 + y^+ y^-, \qquad w_2 = y^+ + y^-, \\ \nonumber
& w_3 = y^+ -y^-, \qquad w_{4} =  1+  y^+ y^-.
\eea
Note that we have use the scaling symmetry of $w_I$'s to set
$w_4 - w_1 =2$. Using the above solution of the constraint \eq{consteq1}
it can be seen 
\be{modconst}
|w_1|^2 + |w_2|^2 - |w_3|^2 - |w_4|^2 = 2Y,
\ee
where $Y = 4y_2^+y_2^-$ 
is related to the K\"{a}hler potential on the moduli space by
\be{realkpot1}
K = -\log Y.
\ee
Now we can parameterize the moduli matrix $u_M$ by
\be{defmodmat1}
u_M = L \tilde U L -L,
\ee
where $\tilde U$ is given by
\be{partu1}
\tilde U = \frac{w_I\bar w_J + \bar w_I w_J}{Y}.
\ee
It can be seen that using \eq{relwystu}, \eq{modconst} and the 
definitions \eq{defmodmat1} and \eq{partu1} the conditions on
$u_M$ given in \eq{condstu} are satisfied. Note that now
we have parameterized the matrix $u_M$ in terms of $y^+$ and $y^-$.
Substituting this parameterization in the entropy function 
\eq{entfns} we obtain  
\be{entfnstu}
F = \frac{\pi}{2} \left[ \frac{|(Q+ \tau P) \cdot w|^2}{\tau_2 Y}
+  \frac{|(Q+ \bar\tau P) \cdot w|^2}{\tau_2 Y}
- \frac{( Q+ \tau P ) \cdot (Q + \bar\tau P)}{ \tau_2} \right],
\ee
where dot product $\cdot$ is with respect to the metric $L$. 
To determine the values of the T and U-moduli 
($y^+, y^-$ respectively ) at the attractor point it is 
sufficient to focus terms on the first two terms in \eq{entfnstu}.
The first two terms are identical except for the exchange of 
$\tau\leftrightarrow \bar\tau$ in the numerator. 
Our strategy for minimizing with respect to $y^+$ and $y^-$ is as follows:
We will first just focus on the first term
\be{tenfstu1}
F_T = \frac{\pi}{2} \left[ \frac{|(Q+ \tau P) \cdot w|^2}{\tau_2 Y} \right],
\ee
and minimize this term with respect to $y^+$ and $y^-$. We will see that 
the attractor values of the moduli $y^+$ and $y^-$ are independent of 
the axion-dilaton moduli $\tau$. Therefore these values of the
attractor moduli minimize the second term in \eq{entfnstu} simultaneously,
since the second term is same as the first term with $\tau\rightarrow
\bar\tau$. Thus to minimize with respect to the $y^+$ and $y^-$ moduli
it is sufficient to focus on \eq{tenfstu1}.

To write out \eq{tenfstu1} in terms of the moduli $y^+$  and $y^-$
it is convenient to define the following variables
\bea{nwinvar1}
Q^1 + Q^4 = N_1, \quad Q^1-Q^4 = W_1, \quad Q^2 + Q^3 = N_2, \qquad Q^2 - Q^3
= W_2, \cr
P^1 + P^4 = \tilde N_1, 
\quad P^1-P^4 = \tilde W_1, \quad P^2 + P^3 = \tilde N_2, \qquad P^2 - P^3
= \tilde W_2.
\eea
We also define the various components of the complex combination $Q+ \tau P$
as follows
\bea{nwvar2}
Q^1 + Q^4 + \tau( P^1 + P^4) =  N_1 + \tau \tilde N_1 = n_1, \quad
Q^1 - Q^4 + \tau( P^1 -P^4) = W_1 + \tau \tilde W_1 = w_1', \cr \nonumber
Q^2 + Q^3 + \tau( P_2 + P_3) =  N_2 + \tau \tilde N_2 = n_2, \quad
Q^2 - Q^3 + \tau( P_2 -P_3) = W_2 + \tau \tilde W_2 = w_2'.
\\
\eea
Using the variables the T-moduli dependent part of the entropy function 
in \eq{entfnstu} is given by
\be{entfnstut}
F_{T} = \frac{\pi}{8\tau_2 y^+_2 y^-_2} 
\left| -n_1 + y^+ y^-  w_1' + y^+ w_2' 
+ y^- n_2 \right|^2,
\ee
where the subscript $T, U$ indicates the term dependent on the 
$T, U$ moduli of the entropy function in \eq{entfnstu}.
Minimizing with respect to $y^+$ and $y^-$ we obtain the following equations
respectively
\bea{mineqyy}
 y^- \bar y^+ w_1' +  w_2' \bar y^+  +
   y^- n_2 - n_1 = 0, \\ \nonumber
 y^+ \bar y^- w_1' +  n_2 \bar y^-
+  y^+ w_2'  - n_1. =0
\eea
Eliminating $y^-$ from the above equations we obtain the following 
quadratic equation for $y^+$ 
\be{eqy+}
( \tilde W_2 W_1 - W_2 \tilde W_1) (y^+ )^2 
+ [ ( \tilde W_2 N_2 - W_2 \tilde N_2) - ( \tilde N_1 W_1 - N_1 \tilde W_1) ]
y^+ - ( \tilde N_1 N_2 - N_1 \tilde N_2 ) = 0.
\ee
Note that in this quadratic equation which determines the moduli $y^+$, the 
axion dilaton dependence completely drops out.
To obtain the solution for  $y^+$ we can simplify the 
discriminant
\bea{discrim}
D &=& \left[ ( \tilde W_2 N_2 - W_2 \tilde N_2) - ( \tilde N_1 W_1 - N_1 \tilde
  W_1)\right]^2
+ 4 (\tilde W_2 W_1 - W_2 \tilde W_1 ) ( \tilde N_1 N_2 - N_1 \tilde N_2) 
\nonumber,\\
 &=&  4 ( ( Q\cdot P)^2 - Q^2 P^2).
\eea
Here we have re-written the charges in terms of $Q$'s and $P$'s using the 
relations in \eq{nwinvar1}. Since we are looking at supersymmetric black holes
we have $Q^2 P^2 - (Q\cdot P)^2>0,   Q^2>0, P^2>0$, 
this implies the discriminant $D$ is
always negative. Thus the solution for $y^+$ is always complex, the
real and imaginary parts of $y^+$ are given by
\bea{soly+}
 y^+_1 & =& - \frac{  ( \tilde W_2 N_2 - W_2 \tilde N_2)
    -( \tilde N_1 W_1 - N_1 \tilde W_1) }{
2 ( \tilde W_2 W_1 - W_2 \tilde W_1 ) }, \\ \nonumber
 y^+_2 &=& 
\frac{\sqrt{ Q^2 P^2 - ( Q\cdot P)^2}}{  ( \tilde W_2 W_1 - W_2 \tilde W_1 )}.
\eea
Note that finally we  have to choose the  solution with $y_2^+>0$, we choose
$\tilde W_2W_1 - W_2\tilde W_1>0$. 
For later purpose it is convenient to write the above solution more 
suggestively as follows.
The electric and magnetic charges are 4-vectors in $SO(2,2; \ZZZ)$.
Let us consider the  antisymmetric combination of these vectors
\be{anstdual}
T^{ij} = Q^i P^j - Q^j P^i,
\ee
where $i, j =1, \cdots 4$,  these 6 components are S-duality invariants.
We can project them into components which transform as a ${\bf{3}}$ of 
the left $SO(2,1;\ZZZ)$ of $SO(2,2;\ZZZ)$ and 
${\bf{3}}$ of the right $SO(2,1;\ZZZ)$ using the self dual and anti-self dual
't Hooft symbols of $SO(2,2)$ respectively. 
From the definition of the 't Hooft symbols in \eq{defseld}
and \eq{adefseld} we see that the solution
\eq{soly+} can be written as
\be{susoly+}
y_1^+ = -\frac{\eta_{1;ij} T^{ij} }{ \eta_{+i;j} T^{ij} }, \quad
y_2^+ = \frac{ \sqrt{ \eta_{a;ij} \eta^a_{\,kl} T^{ij}T^{kl} }}{
\eta_{+;ij} T^{ij} }.
\ee
where $a=1,2,3$. Note that we have used the identity
\eq{a6} to rewrite the invariant $Q^2P^2 - (Q\cdot P)^2$  in the 
above equation.
A similar analysis for the $y^-$ yields the quadratic equation
\bea{eqy-}
( \tilde N_2 W_1 - N_2 \tilde W_1) (y^-)^2
+ \left[ (\tilde N_2 W_2 - N_2 \tilde W_2) + ( \tilde W_1N_1 - \tilde N_1 W_1)
 \right]y^-   + ( \tilde W_2 N_1 - W_2\tilde N_1) =0, \nonumber \\
\eea
here again the axion dilaton dependence drops out.
The discriminant of this quadratic equation is given by
\bea{discrim1}
D&=& \left[ ( \tilde N_2W_2 - N_2 \tilde W_2) + ( \tilde W_1 N_1 - \tilde N_1
  W_1) \right]^2 - 4 ( \tilde N_2 W_1 - N_2 \tilde W_1) ( \tilde W_2 N_1 - W_2
\tilde N_1), \nonumber \\
&=& 4 (Q\cdot P)^2 - Q^2 P^2.
\eea
Thus the real and the imaginary parts of $y^-$ are given by
\bea{soly-}
 y_1^- &=& - \frac{ (\tilde N_2W_2 - N_2 \tilde W_2) + (
  \tilde W_1 N_1 - \tilde N_1 W_1) }{2 ( \tilde N_2 W_1 -N_2 \tilde W_1)}, 
\\ \nonumber
 y_2^- &=&  
\frac{\sqrt{ Q^2 P^2 - (Q\cdot P)^2}}{ \tilde N_2 W_1 - N_2 \tilde W_1}.
\eea
Again we have to choose the solution with $y_2^- >0$ for that we 
have $\tilde N_2 W_1 - N_2 \tilde W_1>0$. Rewriting the solution 
\eq{soly-} in terms of the 't Hooft symbols we obtain
\be{susoly-}
y_1^- = -\frac{\tilde\eta_{1;ij} T^{ij} }{ \tilde\eta_{-;ij} T^{ij} }, 
\qquad
y_2^- = \frac{ \sqrt{ \tilde\eta_{a;ij} \tilde\eta^a_{\,kl} T^{ij}T^{kl} }}{
\tilde\eta_{-;ij} T^{ij} }.
\ee
For later convenience we also write down the real and imaginary 
parts of $-1/y^-$
\bea{susoly-1}
-\left(\frac{1}{y^-} \right)_1 &=& \frac{ (\tilde N_2W_2 - N_2 \tilde W_2) + (
  \tilde W_1 N_1 - \tilde N_1 W_1) }{2 ( \tilde W_2 N_1 -W_2 \tilde N_1)}
=  
\frac{\tilde\eta_{1;ij} T^{ij} }{ \tilde\eta_{+;ij} T^{ij} },
\\ \nonumber
-\left(\frac{1}{y^-} \right)_2& =& 
\frac{\sqrt{ Q^2 P^2 - (Q\cdot P)^2}}{ \tilde W_2 N_1 - W_2 \tilde N_1} =
\frac{ \sqrt{ \tilde\eta_{a;ij} \tilde\eta^a_{\,kl} T^{ij}T^{kl} }}{
\tilde\eta_{+;ij} T^{ij} }.
\eea
The solutions given in \eq{soly+} and \eq{soly-} are independent of the 
axion-dilaton moduli and thus minimize 
both the terms which depend on $y^+$ and $y^-$ in \eq{entfnstu} and 
therefore are the  attractor values for the entropy function.   
The solution for the attractor values of the moduli has been derived earlier
for special charge configurations \cite{Sen:2005iz,Sahoo:2006rp}. 
The analysis given above is 
the complete solution for arbitrary charge configurations, it reduces to the 
solutions found in \cite{Sen:2005iz,Sahoo:2006rp} 
for the respective charge configurations.
The general analysis  enabled us 
to write down the attractor values of the moduli in a  manifestly symmetric
form given in \eq{susoly+}, \eq{susoly-}.

Substituting the attractor 
values of the $T$ and $U$ moduli given in \eq{susoly+} and 
\eq{susoly-} in the entropy function 
\eq{entfnstu} we obtain 
\be{finalent}
F|_{T,U, {\rm{min}}}
= \frac{\pi}{2\tau_2} \left[ (Q + \tau P)\cdot (Q + \bar \tau P) \right].
\ee
We now can further minimize with respect to the axion dilaton moduli
to obtain
\be{soltau}
\tau_1 = -\frac{Q\cdot P}{P^2}, \qquad
\tau_2 = \frac{\sqrt{Q^2 P^2 - (Q\cdot P)^2}}{ P^2}.
\ee
Finally substituting the values of all the vector multiplet moduli
in the entropy function we get the usual Hawking-Bekenstein entropy
\be{hbent}
F|_{{\rm min}}= \pi \sqrt{Q^2P^2 -(Q\cdot P)^2}.
\ee
Note that all the vector multiplets $y^+, y^-, \tau$ are fixed at the 
attractor values, therefore this attractor point has no flat directions.

\subsection{The two derivative entropy function: FHSV model}

Let us call all the vector multiplet moduli other than the 
axion-dilaton moduli as T-moduli. The T-moduli in the FHSV model
parameterizes the coset
\be{vecmod}
{\cal M}_T = \frac{SO(10, 2)}{ SO(10)\times SO(2) },
\ee
To explictly find the values of the T-moduli at the attractor point
we first need to parameterize the $12\times 12$ moduli matrix $u_M$ which 
satisfies the condition
\be{condfshv}
u_M^T = u_M, \qquad u_M^T L u_M = L.
\ee
where $L = {\rm{Dia}}( -1, \cdots, -1, 1, 1)$ is the Lorentzian  metric
with 10 negative 1's and 2 positive 1's  .
From the conditions in \eq{condfshv}
the number of independent variables required to parameterize $u_M$ is
$20$. 
Just as in the STU model we first  introduce $10+2$ complex numbers satisfying
\bea{consteq}
-\sum_{I=1}^{10} w_I^2 +w_{11}^2 + w_{12}^2 = 0,
\eea
together 
with the identification $w_I \sim c w_I$, where $c$  is a complex number.
Note that the  constraint in \eq{consteq}
and the identification of $w$'s upto complex scalings
reduce the number of independent parameters to $20$ which is the required
number of variables to parameterize the moduli space in \eq{vecmod}.
Using the scaling degree of freedom 
 the constraints in \eq{consteq} can be solved by 
introducing the $10$ complex numbers $(y^+, y^-, \vec y)$ where
$\vec y$ is a $8$ dimensional vector. 
These variables are related to the
$12$ $w_I$'s by 
\bea{relwy}
w_I = y_I, \;\;I =1 \cdots 8, &\quad& w_{9} = \frac{1}{\sqrt{2}} ( y^+ -  y^-),
\cr
w_{10} = 1+ \frac{ y^2}{4}, &\quad&
w_{11} = \frac{1}{\sqrt{2}} ( y^+ +  y^-), \cr 
w_{12} = -1 + \frac{y^2}{4}, 
&\quad& y^2 = 2 y^+ y^- + \vec y^2.
\eea
On substituting these values of $w_I$ in \eq{consteq} it is easy to see that
the constraint is satisfied. The above parameterization amounts to 
scaling $w_{10}- w_{12} $ such that its value is constant given by 
$ w_{10}- w_{12}=2$. Using the above solution of  the constraint
\eq{consteq} it can be seen that 
\bea{mconstfs}
-\sum_{I=1}^{10} |w_I|^2 + |w_{11}|^2 + |w_{12}|^2 = 2Y, \\ \nonumber
\hbox{where}\; Y = ({\rm Im} y)^2 = 2 y_2^+ y_2^- - \vec y_2^2. 
\eea
 $Y$ is 
related to the K\"{a}hler potential on the moduli space 
by
\be{relkpot}
K = -\log Y.
\ee
The variables $y$ to parameterize the vector multiplet moduli was introduced in
\cite{Harvey:1996ts,Klemm:2005pd}.
We now parameterize the vector multiplet moduli matrix as 
\be{defmodmat}
u_M = L \tilde U L - L.
\ee
The conditions on the moduli matrix given in \eq{modcond} result in 
the following conditions on $\tilde U$
\be{contu}
\tilde U^T = \tilde U, \qquad
\tilde U L \tilde U - 2 \tilde U =0.
\ee
We can now further parameterize  $\tilde U$ as
\be{partu}
\tilde U = \frac{ w_I \bar w_J + \bar w_I w_J}{ Y},
\ee
here $I =1 \cdots 12$
It is easy to see that given the constraints \eq{consteq} on 
the $w$'s we see that the conditions on $\tilde U$ in \eq{contu} are
satisfied.

Using $\tau$ to parameterize
the axion dilaton moduli as in 
\eq{deftau} and  parameterizing  the moduli matrix $u_M$ in 
terms of $w$ from \eq{defmodmat}, 
\eq{partu} in the entropy function 
\eq{entfns} we can write the entropy function as
\bea{compent}
F &=& \frac{\pi}{2\tau_2 } \left[ ( Q+ \tau P)u_M ( Q+ \bar\tau P) \right], \\ 
\nonumber
& =&  \frac{\pi}{2} \left[  \frac{ | ( Q + \tau P) \cdot w|^2}{\tau_2 Y}  +
\frac{ | ( Q + \bar\tau P) \cdot w|^2}{\tau_2 Y}  
- \frac{( Q+ \tau P)\cdot (Q+ \bar \tau P) }{\tau_2} \right].
\eea
Note that in the above the dot product is taken with the metric  $L$, 
From here it is easy to see that $F$ is both $SL(2,R)$ as well as $SO(10,2;R)$ 
invariant. 
We first  minimize the entropy function
with respect to the T-moduli $(y^+, y^-, \vec y)$ for right moving charges. 
Right moving charges are defined by the condition $Q\cdot Q>0, P\cdot P>0$.
Generic supersymmetric dyons are right moving. 
Details of the minimization procedure are provided in the appendix.
To write down the solution we define the following combination of 
charges
\bea{combchg}
Q^{12} + Q^{10} = N_1, \quad Q^{12} -Q^{10} = W_1, \quad Q^{11} + Q^{9} =
N_2, \quad Q^{11} - Q^{9} = W_2, \\ \nonumber
P^{12} + P^{10} = \tilde N_1, 
\quad P^{12} -P^{10} = \tilde W_1, \quad P^{11} + P^{9} =
\tilde N_2, \quad P^{11} - P^{9} = \tilde W_2.
\eea
We now write down the solution of all the T-moduli: $y^i$, 
$i=1, \cdots 8$  and $y^-$ are determined in
terms of $y^+$ from the equation
\bea{solyiy-}
y^i& =& \sqrt{2} y^+
\frac{Q^i \tilde W_1 - P^i W_1}{ N_2 \tilde W_1 - \tilde N_2 W_1}
+ 2 \frac{\tilde N_1 Q^i - N_2P^i}{ N_2 \tilde W_1 - \tilde N_2 W_1},
\\ \nonumber
y^- &=& y^+ \frac{W_2\tilde W_1 - \tilde W_2  W_1}
{N_2 \tilde W_1 - \tilde N_2 W_1} + \sqrt{2}
\frac{\tilde N_2 W_2 - N_2 \tilde W_2}{N_2 \tilde W_1 - \tilde N_2 W_1},
\eea
$y^+$ is determined from the solution of the following quadratic equation
\bea{solyfy+}
& &A\left( \frac{y^+}{\sqrt{2}}\right)^2 + B \frac{y^+}{\sqrt{2}} 
+ C = 0, \\ \nonumber
& &A = \tilde W_2 W_1 - W_2\tilde W_1 + \sum_{i=1}^8 
\frac{(Q^i \tilde W_1 - P^i W_1)^2 }{ N_2 \tilde W_1 - \tilde N_2 W_1}, \\
\nonumber
& & B = (\tilde W_2 N_2 - W_2 \tilde N_2) -( \tilde N_1 W_1 - N_1 \tilde W_1)
+ \sum_{i=1}^8
\frac{( \tilde N_2 Q^i -P^i N_2)(\tilde W_1 Q^i - P^i W_1) }{
 N_2 \tilde W_1 - \tilde N_2 W_1},
\\ \nonumber
& &C= N_1 \tilde N_2  - \tilde N_1 N_2.
\eea
Note that the coefficients which determine the attractor values  of the moduli
$y^+$ and $y^i$ in \eq{solyiy-} and \eq{solyfy+}
are all functions of the S-duality invariants $Q^I P^J - P^IQ^J$, with
$I, J =1, \cdots 12$. Therefore the attractor values of the T-moduli
do not transform under the S-duality symmetry of the FHSV model.
As a small check on the above solutions for the T-moduli, note that
on setting  the charges $Q^i=0, P=0$ the equations \eq{solyfy+} reduce
to \eq{eqy+} with the $y^+ \rightarrow \sqrt{y^+}$. 
Furthermore from the solutions it is easy to see that under the
scaling $Q^I \rightarrow \lambda Q^I,\; P^I \rightarrow \tilde\lambda P^I$
the attractor values of the T-moduli remain invariant. Thus for independent
scalings of the electric and the magnetic charges the attractor values of the 
T-moduli are invariant. From \eq{solyfy+}, it is see that 
the attractor values of $y^+$ are functions of ratios of 
$B/A, C/A$, thus they are of ${\cal O}(Q^0, P^0 )$ even if the charges 
electric and magnetic charges are scaled with
$\lambda >>1, \tilde \lambda>>1$.
We now substitute the values of the T-moduli into the entropy function 
given in \eq{compent}, this gives
\be{tminentfn}
F = \frac{\pi}{2\tau_2} \left[ (Q + \tau P)\cdot (Q + \bar\tau P)\right].
\ee
Now we proceed with minimizing the above function with respect to the 
axion-dilaton moduli, this results in the following attractor values
\be{soltauf}
\tau_1 = -\frac{Q\cdot P}{P^2}, \qquad
\tau_2 = \frac{\sqrt{Q^2P^2 - (Q\cdot P)^2}}{P^2}.
\ee
Finally substituting all the attractor values of the 
vector multiplet moduli into the 
entropy function we obtain the following Hawking-Bekenstein entropy
\be{finentfs}
F= \pi\sqrt{Q^2P^2 -(Q\cdot P)^2}.
\ee

\subsection{Including the Gauss-Bonnet correction}

In this subsection we include the contribution of the Gauss-Bonnet 
term to the entropy function and evaluate its  contribution to the black hole
entropy. We will retain terms to ${\cal{ O}}(Q^0, P^0)$ terms in charges and 
neglect contributions at ${\cal{O}}(1/Q^2, 1/P^2)$.
The coefficient of the 
 Gauss-Bonnet in both the STU model \eq{cofstugb} and the FHSV model
\eq{cofgb} contains a term proportional to the trace anomaly which 
depends on graviton momentum $p^2$. Since this term does not 
depend on the moduli of the theory the contribution to the 
Gauss-Bonnet term  can be neglected.
Retaining this term in the entropy function formalism just shifts the 
Entropy by a charge independent constant which we do not keep track of.
Therefore we can restrict our attention to the moduli dependent
terms in the coefficient of the Gauss-Bonnet term in \eq{cofstugb}
and \eq{cofgb}.
Consider first a coefficient of the Gauss-Bonnet term to be
\be{excofgb}
\frac{1}{128\pi^2} {\cal F}(\mu, \bar\mu),
\ee
where $\mu$ refers to the axion-dilaton moduli together with the  rest of
the vector multiplet moduli. 
The change in the entropy function due to the above coefficient is
\eq{excofgb} is
\be{chentfn}
\Delta F = \frac{1}{2} {\cal F} (\mu, \bar \mu).
\ee
The total entropy function is now given by
\be{totentfnex}
F = \lambda^2 F^{(2)}( \mu, \bar\mu ) + \frac{1}{2}{\cal  F}
(\mu, \bar\mu),
\ee
where $F^{(2)}$ refers to the two derivative entropy function 
discussed in the previous section. The two derivative term is 
proportional to ${\cal O}(Q^2, P^2)$, this is indicated by the 
coefficient $\lambda^2$ in \eq{totentfnex}.
We can now minimize with respect to the moduli $\mu$ and solve 
for the moduli in powers of $\frac{1}{\lambda^2}$, we denote this 
correction as
\be{coremu}
\mu = \mu^{*} + \frac{1}{\lambda^2} \delta\mu + {\cal O}(1/\lambda^4),
\ee
where $\mu^*$ solves the attractor equations of the 
two derivative entropy function $F^{(2)}$ given by
\be{2derateq}
\left. 
\frac{\partial F^{(2)}( \mu, \bar\mu) }{\partial \mu}
\right|_{\mu =\mu^*} = 0, \qquad
\left. 
 \frac{\partial F^{(2)}( \mu, \bar\mu) }{\partial \bar\mu}
\right|_{\mu = \mu^*} = 0.
\ee 
Since the 
attractor point has no flat directions one can obtain a unique solution
for $\delta \mu$. Substituting the solution back into $F$ given by
\eq{totentfnex} and using \eq{2derateq}  it is easy to see that to 
${\cal{O}}(Q^0, P^0)$ in charges  the entropy is given by
\be{valentmin}
F = \lambda^2 F^{(2)} (\mu^*, \bar\mu^*) + 
\frac{1}{2} {\cal  F}(\mu^*, \bar\mu^*) + {\cal O}(1/\lambda^2).
\ee
Thus to evaluate the correction to entropy due to the Gauss-Bonnet term
to ${\cal O}(Q^0, P^0)$ in charges  it is sufficient to 
evaluate its coefficient at the attractor values of the moduli.

\vspace{.5cm}
\noindent
{\emph{STU model}}
\vspace{.5cm}

To evaluate the correction to the entropy including the Gauss-Bonnet 
term to ${\cal O}(Q^0, P^0)$ in charges we substitute the values of the 
vector multiplet at the attractor point given in 
\eq{susoly+}, \eq{susoly-} and  \eq{soltau} into the 
coefficient of the Gauss-Bonnet term and using \eq{valentmin}. We obtain
\bea{stusolentfn}
 F &=& \pi\sqrt{Q^2P^2 - (Q\cdot P)^2}  \\ \nonumber 
& &-\ln\left[\vartheta_2 (\tau)^4 \vartheta_2 (-\bar\tau)^4(\tau_2)^2 
\right]|_*
-\ln\left[\vartheta_2 (y^+)^4 \vartheta_2 (-\bar y^+)^4(\tau_2)^2 
\right]|_* \\ \nonumber
& &-\ln\left[\vartheta_2(-1/y^-)^4 \vartheta_2 (1/\bar y^-)^4((-1/y^-)_2)^2 
\right]|_* + {\cal{O}}(1/Q^0, 1/P^0).
\eea
Here the subscript $*$ refers to the fact that we have substitute the
attractor
values for the moduli given in \eq{susoly+}, \eq{susoly-} and  \eq{soltau}.
Note that using a modular transformation we have written the
$\vartheta_4(y^-)$ in terms of $\vartheta_2(-1/y^-)$ in the 
coefficient of the Gauss-Bonnet term. 

From the expressions for the moduli in \eq{susoly+}, \eq{susoly-} and  
\eq{soltau} and the fact that the leading term in the entropy is 
invariant under triality,  it is easy to see that
the corrected entropy given in \eq{stusolentfn} is invariant under the
following triality symmetries.
\be{entrisym1}
{\cal T}_1: W_1\leftrightarrow \tilde N_2,    \qquad ;
W_2\leftrightarrow -\tilde N_1.
\ee
Under this exchange of charges the attractor
value of the moduli transform as
\be{atexsy1}
\tau_*\leftrightarrow y^+_*,
\quad  y^-_*\; \hbox{invariant}.
\ee
Similarly  under the exchange of the charges
\be{entrisym2}
{\cal T}_2: W_1 \leftrightarrow N_1, \qquad ;\tilde 
 W_1 \leftrightarrow \tilde N_1, 
\ee
the attractor values of the  moduli transform as
\be{atexsy2}
y^+\leftrightarrow_* -\frac{1}{y^-_*},
\qquad   \tau\; \hbox{invariant}.
\ee
Finally under the exchange of charges
\be{entrisym3}
{\cal T}_3: \tilde N_2 \leftrightarrow N_1\qquad ;
\tilde W_1\rightarrow -W_2,
\ee
the attractor values of the moduli transform as
\be{atexsy3}
\tau_*\rightarrow -\frac{1}{y^-_*},  \qquad
y^+\; \hbox{invariant}.
\ee

In the next section we write down a partition function which for dyons 
in the STU model which captures the subleading coefficient obtained by 
considering the Gauss-Bonnet term. The partition function manifestly has 
all the symmetries of the model.

\vspace{.5cm}
\noindent
{\emph{The FHSV model}}
\vspace{.5cm}

To obtain the entropy of the dyonic black hole in the FHSV model with the
 Gauss-Bonnet  term to ${\cal O}(Q^0,P^0)$ in charges we substitute the 
 attractor values of the the moduli given in \eq{solyiy-}, \eq{solyfy+} in
 \eq{valentmin}. The coefficient of the Gauss-Bonnet term is obtained from
 \eq{threshfs}. Performing this we obtain
\bea{fshvf1}
F &= &\pi \sqrt{Q^2P^2 -(Q\cdot P)^2} - 
\ln\left( \eta(2\tau)^{12} \eta( -2\bar \tau)^{12} \tau_2^{6} \right)|_* \\ \nonumber
& & -\frac{1}{2}\ln\left( \Phi_{\rm{BE}}(y) \Phi_{\rm{BE}}
(-\bar y) ( 2y_2^+ y_2^-  -\vec y^2)^4 \right)|_* + {\cal{O}}( 1/Q^2, 1/P^2).
\eea
In this model there is a scaling limit of the charges in which the contribution of
entropy from the 
the T-moduli becomes negligible compared that of the axion-dilaton moduli.
Let us consider the following scaling of the charges 
\bea{scalimit}
& &Q^I\rightarrow \lambda Q^I,\qquad P^I\rightarrow \lambda'P^I  \quad {\hbox{with}}\;
\lambda>>1, \; \lambda'>>1 \\ \nonumber
& & {\hbox{ and}} \;  \lambda>>\lambda'. 
\eea
From the solution of the T-moduli given in \eq{solyiy-}, \eq{solyfy+}, it is 
easy to see that in this scaling limit the T-moduli are of order 1, 
${\cal O}(\lambda^0, \lambda^{'0})$. The axion dilaton moduli on the other
hand scales as 
\be{scaleaxdi}
\tau_1\rightarrow \frac{\lambda}{\lambda'} \tau_1, \qquad
\tau_2\rightarrow \frac{\lambda}{\lambda'} \tau_2.
\ee
Thus in \eq{fshvf1} we can neglect the contribution of the T-moduli and retain 
the leading contribution from the axion-dilaton moduli. Therefore  \eq{fshvf1} 
reduces to 
\bea{scalentfn}
F &=& \pi \sqrt{Q^2P^2 -(Q\cdot P)^2} + 2\pi \tau_2|_* - 6\ln \tau_2|_ *+ 
{\cal{ O}}( Q^{0}, P^0)  \\ \nonumber
& =& \pi \sqrt{Q^2P^2 -(Q\cdot P)^2} + 2 \pi \frac{ \sqrt{Q^2P^2 -(Q\cdot P)^2} }{P^2}
-6 \ln \left( \frac{ \sqrt{Q^2P^2 -(Q\cdot P)^2} }{P^2}\right) \\ \nonumber
& & + {\cal{ O}}( Q^{0}, P^0).
\eea
In section 4. we will write down an approximate dyon  partition function involving a 
Siegel modular form of weight 4 which captures the degeneracy of the dyons in the
FHSV model in the limit when the electric charges are much larger than the 
magnetic charges as in \eq{scalimit}. Note that in this limit though we loose most
of the details of the correction to the entropy from the Gauss-Bonnet term, 
the information of the weight of the $SL(2, \ZZZ)$ modular form which captures 
the axion-dilaton dependence is present in the coefficient of the logarithm
in \eq{scalentfn}.

\section{Partition function for the STU model}

In this section we propose a partition function for the dyons 
of the  STU model. The partition function is given in terms
of product of three Siegel modular forms of weight zero. 
For the purposes of extracting the degeneracy from the 
partition function we need to define the charge bilinears
which occur in the Fourier expansion of the partition function.
From the consideration of the entropy function in the pervious 
section we have seen that that the following charge bilinears
characterize the black hole entropy of the dyons to the next leading
order. There are three sets of charge bilinears, each set is invariant
with respect to any  two of the three $\Gamma(2)$ symmetries of the STU
model. The T-duality and U-duality invariants are given by
\be{tdualinv}
M_{\sigma^1} = \frac{1}{3} Q\cdot Q , 
\qquad M_{\rho^1}  =  \frac{1}{3} P\cdot P, \qquad
M_{v^2} = \frac{1}{3}  Q\cdot P. \\ \nonumber
\ee
The S-duality invariants are given by following anti-symmetric second 
rank tensor $T^{ij}$ of $SO(2,2)$
\be{sdualinv}
T^{ij} = Q^iP^j - Q^j P^i
\ee
where $i, j = 1, \cdots 4$. From the above 6 components of the 
S-duality invariant 
the  charge bilinears  we can project into the self-dual or the 
anti-self dual combinations which are further invariant under 
either T-duality or the U-duality symmetries. This projection 
is done by the 't Hooft symbols of $SO(2,2)$.   They are given by 
\bea{sdualcomb}
M_{\sigma^2} &=&\frac{1}{3} \eta_{-;ij}  T^{ij}, \qquad 
M_{\rho^2} = \frac{1}{3}  \eta_{+;ij }  T^{ij}, \qquad
M_{v^2} = \frac{1}{3}\eta_{1;ij} T^{ij}, \\ \nonumber
M_{\sigma^3} &=& \frac{1}{3} \tilde \eta_{-;ij} T^{ij}, \qquad
M_{\rho^3} =\frac{1}{3}  \tilde \eta_{+;ij}  T^{ij}, \qquad
M_{v^3} = -\frac{1}{3} \tilde \eta_{1;ij} T^{ij}.   
\eea
Here $\eta_{a, ij}$ and $\tilde \eta_{a, ij}$ are the 't Hooft symbols which 
decomposes the second rank anti-symmetric tensor $T^{ij}$ of $SO(2,2)$ and 
as $({\bf{3}},0)$ and $(0,{\bf{3}})$ of $SU(1,1)\times SU(1,1)$. $a, =1, \cdots 3$.
These 't Hooft symbols are defined in  appendix A. 

We now propose the partition function for this model:
The partition function involves the product of inverses of three $Sp(2,\ZZZ)$
modular forms of weight zero and is given by
\bea{stupart} 
d(M_{\rho}, M_{\sigma}, M_v) 
&=&\prod_{\alpha =1}^3  I_{\alpha} (M_{\rho^\alpha}, M_{\sigma^\alpha}, M_{v^\alpha}),
\\ \nonumber
 I_{\alpha} (M_{\rho^\alpha}, M_{\sigma^\alpha}, M_{v^\alpha}) &=&
\frac{K}{4}\exp{(i\pi  M_v)}\int _{{\cal{C}^\alpha}}  
d\tilde\rho^\alpha d\tilde \sigma^\alpha d\tilde v^\alpha \frac{1}{\Phit_0(\tilde\rho^\alpha,\tilde\sigma^\alpha,\tilde v^\alpha)} \\ \nonumber
& \times& \exp\left(-2\pi i \left[ \frac{M_{\rho^\alpha}}{2} \tilde\rho^\alpha
+\frac{M_{\sigma^\alpha} }{2}\tilde\sigma^\alpha  + M_{v^\alpha} \tilde v
\right] \right).
\eea
The $Sp(2,\ZZZ)$ modular form $\Phit_0(\tilde\rho, \tilde\sigma, \tilde v)$ 
which occurs in the partition function is given 
by
\be{defphi0}
\Phit_0 = \Phit_2 \sqrt{ \frac{\Phit_2'}{\Phit_6} }.
\ee
Here $\Phit_6$ is the Siegel modular form of weight 6 under the subgroup
$\tilde G$ of $Sp(2,\ZZZ)$ which captures the degeneracy of dyons in 
the $\ZZZ_2$ CHL model \cite{Jatkar:2005bh,David:2006ji}
 $\Phit_2$ is the Siegel modular form 
of weight $2$ under the same subgroup $\tilde G$ of $Sp(2,\ZZZ)$ which 
captures the degeneracy of dyons in the $\ZZZ_2$ orbifold of type IIB theory
\cite{David:2006ru}.
Finally $\Phit_2'$ is  the modular form of weight $2$ related to $\Phit$ by
\be{relphi2s}
\Phit_2'(\tilde\rho,\tilde,\sigma, \tilde v) = 
\Phit_2(\frac{\tilde\sigma}{2}, 2\tilde\rho, \tilde v).
\ee
From property 1 of $\Phit_2'$ listed in the appendix C, we see that
$\Phit_2'$ is also a Siegel modular form of weight 2 under the subgroup
$\tilde G$ of $Sp(2, \ZZZ)$.  
It is clear from the definition of $\Phit_0$ that it is a modular form
of weight 0, under the same subgroup $\tilde G$ of $Sp(2, \ZZZ)$. 
All the modular forms given in \eq{defphi0} are explicitly constructed in the 
appendix C in their product representation from appropriate 
threshold integrals. 
In \eq{stupart} the three dimensional contour ${\cal{C}}^\alpha$ is given by
\bea{phi0cont}
{\rm Im}\, \tilde\rho^\alpha = M_1^\alpha, \; 
{\rm Im}\, \tilde\sigma^\alpha = M_2^\alpha, \;
{\rm Im}\, \tilde v^\alpha = M_3^\alpha,\\ \nonumber
0\leq {\rm Re} \,\tilde\rho^\alpha \leq 2, \;
0\leq {\rm Re}\, \tilde\sigma^\alpha \leq 2, \;
0\leq {\rm Re} \,\tilde v^\alpha \leq 1.
\eea
The constant $K = -2^{-10}$. 
From the Fourier expansion of the 
partition function it can be shown that 
the  degeneracy formula in \eq{stupart} is 
valid with 
\be{validdeg}
M_{\rho^\alpha} \in \ZZZ, \quad
M_{\sigma^\alpha} \in \ZZZ, \quad
M_{v^\alpha} \in \ZZZ.
\ee
We now list the properties of $\Phit_0$ which are proved in the
appendix C.

\vspace{.5cm}
\noindent
{\emph{Properties of $\Phit_0$}}
\vspace{.5cm}
\begin{enumerate}
\item
$\Phit_0$ is a modular form of weight 0 under the subgroup
$\tilde G$ os $Sp(2, \ZZZ)$. 
\item
$\Phit_0(\tilde\rho,\tilde\sigma, v)$ is an analytic function with second order zeros at
\bea{zphi0}
& &n_2(\tilde\sigma,\tilde\rho -\tilde v^2) + b \tilde v + n_1\tilde\sigma -\tilde\rho m_1 + m_2 =0, \\ \nonumber
& & {\hbox{for}}\; m_1n_1+ m_2 n_2 + \frac{b^2}{4} = \frac{1}{4},\; m_1 \in 2\ZZZ, n_1\in\ZZZ,
b\in 2\ZZZ+1 , m_2, n_2 \in \ZZZ.
\eea
It has second order poles at 
\bea{pphi0}
& & 
n_2(\tilde\sigma,\tilde\rho -\tilde v^2) + b \tilde v + n_1\tilde\sigma -\tilde\rho m_1 + m_2 =0, \\ \nonumber
& & {\hbox{for}}\; m_1n_1+ m_2 n_2 + \frac{b^2}{4} = \frac{1}{4},\; m_1 \in 2\ZZZ+1, n_1\in\ZZZ,
b\in 2\ZZZ+1 , m_2, n_2 \in \ZZZ.
\eea
and first order poles at 
\bea{p1phi0}
& & n_2(\tilde\sigma,\tilde\rho -\tilde v^2) + b \tilde v + n_1\tilde\sigma -\tilde\rho m_1 + m_2 =0, \\ \nonumber
& & {\hbox{for}}\; m_1n_1+ m_2 n_2 + \frac{b^2}{4} = \frac{1}{4},\; m_1 \in 2\ZZZ, n_1\in\ZZZ
+\frac{1}{2},
b\in 2\ZZZ+1 , m_2, n_2 \in \ZZZ.
\eea
\item
$\Phit_0$ is invariant under the following $Sp(2,\ZZZ)$ transformations
\bea{sp2zt}
\Phi_0(\tilde\rho', \tilde\sigma', \tilde v') = 
\Phi_0(\tilde\rho,\tilde\sigma,\tilde v), 
\\ 
\label{tt1}
\hbox{for}\; \tilde\rho'= \tilde\rho + 4\tilde\sigma + 4 \tilde v -2, \quad
\tilde\sigma' = \tilde\sigma, \quad \tilde v' = 2\tilde\sigma + \tilde v, 
\\ 
\label{tt2}
\hbox{and} \; \tilde\rho'= \tilde\rho, \quad
\tilde\sigma' = 4\tilde\rho +\tilde\sigma+ 4\tilde v -2 , 
\quad \tilde v' = 2\tilde\sigma + \tilde v.
\eea
This property is due to the fact that $\Phit_0$ is invariant under the 
subgroup of $\tilde H$ defined in \eq{subgroup}. The transformation 
\eq{tt1} corresponds to the choice $a= -1, b=2, c=0, d=-1$ in \eq{sghtrans}
and  \eq{tt2} corresponds to the choice $a =1, b =0, c=-2, d=1$ in
\eq{sghtrans}. The above $Sp(2, \ZZZ)$ transformations are 
essentially the generators of $\Gamma(2)$. 
Note that  $\Phit_0$ is invariant under the
subgroup \eq{subgroup} which is $\Gamma_0(2)$ which contains the 
group $\Gamma(2)$.  
\item
 The expansion of $1/\Phit_0$ in terms of Fourier coefficients  given by
\be{fexp}
\frac{K}{\Phit_0(\tilde\rho,\tilde\sigma, \tilde v)} = 
\sum_{\stackrel{m, n, p}{m\geq -1/2, n \geq 1/2}}
e^{2\pi i( m \tilde\rho + n\tilde\sigma + p\tilde v)}
g(m , n, p),
\ee
with $g(m, n, p)$ being integers,  $p\in \ZZZ$ while $m, n\in \ZZZ/2$ 
\item
For small $\tilde\rho\tilde\sigma - \tilde v^2 + \tilde v$  $\Phit_0$ factorizes as
\be{facphi0}
\Phit_0(\tilde\rho,\tilde\sigma, \tilde v) = 
4\pi^2 ( 2v - \rho - \sigma)^2 v^2 f^{(0)}(\rho) f^{(0)}(\sigma)
+ {\cal O} (v^2),
\ee
where 
\be{deff2}
f^{(0)} (\rho) = \vartheta_2(\rho)^4.
\ee
The variables $(\rho, \sigma, v)$ and  
$(\tilde\rho, \tilde\sigma, \tilde v)$ are related by the 
$Sp(2, \ZZZ)$ transformation 
\be{chgvar}
\rho= \frac{\tilde\rho\tilde\sigma  - \tilde v^2}{\tilde \sigma}, \qquad
\sigma = \frac{\tilde\rho\tilde\sigma - (\tilde v -1)^2}{\tilde \sigma}, 
\qquad
v = \frac{\tilde\rho\tilde\sigma - \tilde v^2 + \tilde v}{\tilde \sigma},
\ee
or the inverse relations
\be{chgvar2}
\tilde\rho = \frac{v^2 - \rho\sigma}{2 v - \rho -\sigma}, \qquad
\tilde\sigma = \frac{1}{2 v -\rho - \sigma}, 
\qquad
\tilde v = \frac{v -\rho}{2 v - \rho -\sigma}.
\ee
\end{enumerate}

We now perform several check on the proposal \eq{stupart}, we show
that it has all the duality symmetries of the STU model and then
show that for large charges the statistical entropy obtained from 
the partition function reproduces the degeneracy of dyonic
black holes in the STU model to the first subleading order in 
charges.

\subsection{Consistency checks}

In this section we will subject our proposal given in \eq{stupart} to various
consistency checks. We first verify that the integrand in \eq{stupart}  has 
STU triality symmetry and then show that it has
$\Gamma(2)_S\times \Gamma(2)_T\times \Gamma(2)_U$ 
symmetry.

\vspace{.5cm}
\noindent
{\emph{STU triality symmetry}}
\vspace{.5cm}

From the definitions of the charge bilinears it is easy to show the following
transformation properties:
Under
\bea{dtrail1}
{\cal T}_1: \;W_1\leftrightarrow \tilde N_2,  \qquad
W_2\leftrightarrow -\tilde N_1,
\eea
the following charge bilinears transform as 
\bea{trail1}
M_{\rho^1} \leftrightarrow M_{\rho^2}, \quad
\quad M_{\sigma^1}\leftrightarrow M_{\sigma^1},
\quad M_{v^1} \leftrightarrow M_{v^2},  
\eea
while the charge bilinears $M_{\rho^3}, M_{\sigma^3}, M_{v^3}$ 
remains invariant. It is now easy to see that 
under this  transformation ${\cal T}_1$the integrand in the partition 
function \eq{stupart} is manifestly invariant. 
Similarly under
\bea{dtrail2}
{\cal T}_2: \;W_1\leftrightarrow  N_1, \qquad 
\tilde W_1\leftrightarrow \tilde N_1,
\eea
that is the exchange of momentum and winding on  the first circle, 
the following charge bilinears transform as 
\bea{trail2}
M_{\rho^2} \leftrightarrow M_{\rho^3}, \quad
 M_{\sigma^2}\leftrightarrow M_{\sigma^3}, \quad
 M_{v^2} \rightarrow M_{v^3},  
\eea
while the charge bilinears $M_{\rho^1}, M_{\sigma^1}, M_{v^1}$ 
remains invariant. Thus it is easy to see that
the integrand in the partition 
function \eq{stupart} is manifestly 
invariant under the transformation ${\cal T}_2$. 
Finally the transformation
\bea{dtrail3}
{\cal T}_3: \;\tilde N_2\leftrightarrow  N_1, \qquad
 \tilde W_1\leftrightarrow -W_2, 
\eea 
the following charge bilinears transform as 
\bea{trail3}
M_{\rho^1} \leftrightarrow M_{\rho^3}, \quad
 M_{\sigma^1}\leftrightarrow M_{\sigma^3},
\quad M_{v^1} \rightarrow M_{v^3},  
\eea
while the charge bilinears $M_{\rho^2}, M_{\sigma^2}, M_{v^2}$ 
remains invariant. Thus it is easy to see that
the integrand in the partition 
function \eq{stupart} is manifestly 
invariant under the transformation ${\cal T}_3$. 
Therefore ignoring considerations of the contour ${\cal{C}^\alpha}$ the 
degeneracy is invariant under the 
triality symmetries ${\cal T}_1, {\cal T}_2, {\cal T}_3$.

\vspace{.5cm}
\noindent
{\emph{$\Gamma(2)_S\times \Gamma(2)_T\times \Gamma(2)_U$ symmetry}}
\vspace{.5cm}

We first look at $\Gamma(2)_S$ action, which acts on the 
electric and magnetic charges as
\be{sdualtra}
Q\rightarrow Q' = a Q+ b P, \quad
P \rightarrow P' = cQ + d P , \quad
\left(
\begin{array}{cc}
a & b \\
c & d
\end{array}
\right) \in \Gamma(2).
\ee
The generators of $\Gamma(2)$ are given by the following matrices
\be{gengamma2}
M_{\infty} = \left(
\begin{array}{cc}
-1 & 2 \\
0 & -1
\end{array} \right), \qquad
M_{1} = \left(
\begin{array}{cc}
1 & 0 \\
-2 & 1
\end{array}
\right).
\ee
Using \eq{sdualtra} with the generator $M_\infty$ we find that the 
T-duality invariants $M_{\rho^1}, M_{\sigma^1}, M_{v^1}$ transform as
\be{tstrans}
\left(
\begin{array}{c}
M_{\sigma^{1}}'\\
M_{\rho^{1}}'\\
M_{v^{1}}'
\end{array}
\right)
=
\left(\begin{array}{ccc}
1 & 4 & -4 \\
0 & 1 & 0 \\
0 & -2 & 1 
\end{array}\right)
\left(
\begin{array}{c}
M_{\sigma^1}\\
M_{\rho^1}\\
M_{v^1}
\end{array}
\right).
\ee 
Now following \cite{Jatkar:2005bh} let us define
\be{defgamm2om}
\tilde\Omega^{1' }\equiv
\left(\begin{array}{cc}
\tilde\rho^{1'} &\tilde v^{1'}  \\
\tilde v^{1'}  & \tilde\sigma^{1\prime}
\end{array}
\right)
= (\tilde A\tilde\Omega^1 + \tilde B) ( \tilde C\tilde\Omega^1  + \tilde D)^{-1}, \quad
\left(
\begin{array}{cc}
\tilde A & \tilde B \\
\tilde C & \tilde D
\end{array}
\right)
= \left(
\begin{array}{cccc}
a & -b & b &0 \\
-c & d & 0 & c \\
0 & 0& d & c \\
0 & 0 & b & a 
\end{array}
\right).
\ee
For the matrix $M_\infty$ we obtain
\bea{minftytr}
\tilde\rho^{1' }&=& \tilde\rho^1 + 4\tilde\sigma^1 + 4 \tilde v^1 -2, \cr
\tilde\sigma^{1'} &=& \tilde\sigma^1, \cr
\tilde v^{1'} &=& 2\tilde\sigma^1 + \tilde v^1.
\eea
Using the transformations \eq{tstrans} and \eq{minftytr} we
obtain the following transformations
\bea{mestrans}
& & \exp\left[2\pi i \left( \frac{M_{\rho^1}'}{2} \rho^{1'} +
 \frac{M_{\sigma^1}'}{2} \sigma^{1'} + M_{v^1}' v^{1'} \right)\right] \\ \nonumber
 & & \;\;\;\;\;\;\;\;\; \rightarrow
 \exp\left[2\pi i \left( \frac{M_{\rho^1}}{2} \rho^{1} +
 \frac{M_{\sigma^1}}{2} \sigma^{1} + M_{v^1}v^{1} \right)\right]
 \exp{(2\pi i M_{\rho^1})}, 
 \\ \nonumber
& &  \exp{(i\pi M_{v^1}')} 
 \rightarrow \exp{(-2\pi  i M_{\rho^1})} \exp{(i\pi M_{v^1})}.
 \eea
 Furthermore from \eq{minftytr} and  \eq{sp2zt} we see that 
\be{invarphi}
\Phit_0(\tilde\rho^{1'}, \tilde\sigma^{1'}, \tilde v^{1'} ) = \Phit_0
(\tilde\rho^1, \tilde\sigma^1,
\tilde v^1).
\ee 
 Finally one can show that 
\be{mesutrans}
d\tilde\rho^{1'}d \tilde\sigma^{1'}d \tilde v^{1'} =d\tilde\rho^1d\tilde\sigma^1d
\tilde v^1.
\ee
Combining \eq{mestrans}, \eq{mesutrans} and \eq{invarphi} we see that the 
the integrand in $I_1$ is invariant under $M_\infty$. Since the transformations in
\eq{tstrans} are S-duality transformations they leave the 
leave the the remaining  charge bilinears in \eq{sdualcomb} invariant. 
Therefore  the integrand in the partition function \eq{stupart} remains
invariant under the S-duality action of $M_\infty$. 
Using the same argument one can show that the integrand in 
\eq{stupart} is invariant under the S-duality action of $M_1$ under which 
the charge bilinears transform as
\be{tstrans2}
\left(
\begin{array}{c}
M_{\sigma^{1}}'\\
M_{\rho^{1}}'\\
M_{v^{1}}'
\end{array}
\right)
=
\left(\begin{array}{ccc}
1 & 0 & 0 \\
4 & 1 & -4 \\
-2 & 0 & 1 
\end{array}\right)
\left(
\begin{array}{c}
M_{\sigma^{1}}\\
M_{\rho^{1}}\\
M_{v^{1}}
\end{array}
\right),
\ee 
and the $Sp(2,\ZZZ)$ variables transform as the equations in the third line
of \eq{sp2zt}.
We can thus conclude that the integrand is invariant under $\Gamma(2)_S$.
In fact since the form $\Phit_0$ is invariant under 
$\Gamma_0(2)$ (see property 1 of $\Phit_0$ in appendix C.), 
the partition function is invariant under this larger symmetry.

The action of the two generators $M_\infty$ and 
$M_1$ of $\Gamma(2)_T$ on the charge bilinears $M_{\rho^2}, M_{\sigma^2}, 
M_{v^2}$ is given by the the same equations \eq{tstrans} and \eq{tstrans2}
with the $1$ replaced by $2$. 
To show that the integrand is 
invariant under $\Gamma(2)_T$ is is sufficient to use
the fact that the integrand is invariant under the 
triality symmetry ${\cal{T}}_1$.
From the action of ${\cal T}_1$ on the charge bilinears 
given in \eq{dtrail1}, \eq{trail1} we have the following 
relation between the action of $\Gamma(2)_T$ and $\Gamma(2)_S$
\be{relts}
\Gamma(2)_T = {\cal T}_1 \Gamma(2)_S {\cal T}_1.
\ee
Since the integrand is invariant under $\Gamma(2)_S$ and the triality symmetry
${\cal T}_1$ we see that the $\Gamma(2)_S$ is a symmetry of the integrand.
Finally the action of the two generators $M_\infty$ and 
$M_1$ of $\Gamma(2)_U$ on the charge bilinears $M_{\rho^2}, M_{\sigma^2}, 
M_{v^2}$ is given by the the same equations \eq{tstrans} and \eq{tstrans2}
with the $1$ replaced by $3$. Using \eq{dtrail3} and \eq{trail3} we have
\be{relts1}
\Gamma(2)_U = {\cal T}_3  \Gamma(2)_S {\cal T}_3.
\ee
Since the integrand is invariant under $\Gamma(2)_S$ and the triality symmetry
${\cal T}_3$ we see that the $\Gamma(2)_S$ is a symmetry of the integrand.
Thus we conclude that ignoring considerations of the contour
the degeneracy given by  \eq{stupart} is invariant under 
$\Gamma(2)_S\times \Gamma(2)_T\times\Gamma(2)_U$.

\vspace{.5cm}
\noindent
{\emph{Integrality of the $d(M_\rho, M_{\sigma}, M_v)$}}
\vspace{.5cm}

From property \eq{fexp} of $\Phit_0$ that Fourier coefficients $d(M_\rho,
M_\sigma, M_v)$ are all integers. From our definition of the charge bilinears
in \eq{sdualcomb}  and from the moding of the Fourier expansions in 
\eq{fexp} we see that 
\be{chgquant}
M_{\rho^\alpha} = \ZZZ, \qquad
M_{\sigma^\alpha} = \ZZZ, \qquad
M_{v^\alpha} = \ZZZ.
\ee
Thus the class of dyons the partition function given in \eq{stupart} is
applicable has the above quantization conditions.

\subsection{Statistical entropy of the STU model}

In this section we obtain the asymptotic degeneracies for dyons 
for large charges $M_{\rho^\alpha}, M_{\sigma^\alpha}, M_{v^\alpha}>>0$
with $Q^2P^2-(Q\cdot P)^2 >>0$. 
We begin with the formula \eq{stupart} for the degeneracy of dyons in the
STU model.  The degeneracy is obtained as a product of three integrals 
given by
\bea{statint}
 I_{\alpha} (M_{\rho^\alpha}, M_{\sigma^\alpha}, M_{v^\alpha}) &=&
\frac{K}{4}\exp{(i\pi  M_{v^\alpha})}\int _{{\cal{C}^\alpha}}  
d\tilde\rho^\alpha d\tilde \sigma^\alpha d\tilde v^\alpha \frac{1}
{\Phit_0(\tilde\rho^\alpha,\tilde\sigma^\alpha,\tilde v^\alpha)} \\ \nonumber
& \times& \exp\left[-2\pi i \left( \frac{M_{\rho^\alpha}}{2} \tilde\rho^\alpha
+\frac{M_{\sigma^\alpha} }{2}\tilde\sigma^\alpha  + M_{v^\alpha} \tilde v
\right) \right].
\eea
This formula is identical in form to equation (3.29) of \cite{David:2006yn}
with the substitution 
$Q_e^2 \rightarrow M_{\sigma^\alpha},\; Q_m^2 \rightarrow
M_{\rho^\alpha}, \;Q_e\cdot Q_m \rightarrow M_{v^\alpha}$. 
Following \cite{David:2006yn} one can show that the dominant contribution
to this integral comes form the residue at the pole at
\be{dompole}
\tilde\rho^{\alpha} \tilde\sigma^\alpha - (\tilde v^{\alpha})^2 + 
\tilde v^\alpha = 0.
\ee
The behavior of $\Phit_0$ near this zero is given by 
\eq{facphi0}, is identical to the corresponding relation (4.17) in 
\cite{David:2006yn} with $k\rightarrow 0$ and 
$f^{(k)}(\rho)  \rightarrow f^{(0)}(\rho)$
given in \eq{deff2}. Thus following an analysis identical
to that in \cite{David:2006yn} we can conclude that for large charges the 
contribution to the statistical entropy form the integral $I_{\alpha}$ 
defined as the log of the contribution of the degeneracy 
 $I_{\alpha} (M_{\rho^\alpha}, M_{\sigma^\alpha}, M_{v^\alpha})$ is obtained 
by minimizing the statistical entropy function 
\bea{entstu}
-\tilde\Gamma_B^\alpha (\vec\gamma^\alpha) &=&
\frac{\pi}{2\gamma_2^\alpha} 
\left( M_{\sigma^\alpha} + 2\gamma_1^\alpha M_{v} + \tau^\gamma\bar\tau^\gamma
 M_{\rho^\alpha} \right)\\ \nonumber
& & -\ln\left[ \vartheta_2(\gamma^\alpha)^4 \vartheta_2(-\bar\gamma^\alpha)^4
(2\gamma_2^\alpha)^2 \right] + \hbox{constant} + {\cal O}(1/Q^2),
\eea
with respect to the real and imaginary parts of $\gamma^\alpha$.
To order ${\cal O}(1/Q^2)$ it is sufficient to obtain the value 
of $\tau^\alpha $ at the minimum by minimizing the ${\cal O}(Q^2)$ term in 
\eq{entstu}. This is given by
\bea{mintau}
\gamma_1^\alpha|_*
 &=& -\frac{M_{v^{\alpha}}}{M_{\rho^\alpha}}, \\ \nonumber
\gamma_2^\alpha|_{*}  &=& \frac{
\sqrt{ M_{\rho^\alpha} M_{\sigma^\alpha} - M_{v^\alpha}^2 }}
{M_{\rho^\alpha}}, \\ \nonumber
&=& \frac{ \sqrt{Q^2P^2 - (Q\cdot P)^2} }{3 M_{\rho^\alpha}}.
\eea
We can
substituting these values in the statistical entropy function \eq{entstu}
to obtain the value of the statistical entropy from 
$I_\alpha$ to ${\cal O}(1/Q^2)$, this is given by
\bea{valminsent}
-\tilde\Gamma_B(\vec\gamma^\alpha)|_*
&=& \pi 
\sqrt{  M_{\rho^\alpha} M_{\sigma^\alpha} - M_{v^\alpha}^2 } \\ \nonumber
 & & 
-\left. \ln\left[ \vartheta_2(\gamma^\alpha)^4 
\vartheta_2(-\bar\gamma^\alpha )^4
(2\gamma_2^\alpha)^2 \right]\right|_* + \hbox{constant} + {\cal O}(1/Q^2),
\\ \nonumber
&=& \frac{\pi}{3} 
\sqrt{Q^2P^2 - (Q\cdot P)^2 } \\ \nonumber
 & & 
-\left. \ln\left[ \vartheta_2(\gamma^\alpha)^4 
\vartheta_2(-\bar\gamma^\alpha )^4
(2\gamma_2^\alpha)^2 \right]\right|_* + \hbox{constant} + {\cal O}(1/Q^2).
\eea
The total statistical entropy is then given by 
\be{stattot}
-\tilde\Gamma_B(\vec\gamma^1, \vec\gamma^2, \vec\gamma^3) 
= -\sum_{\alpha =1}^3 \tilde\Gamma_B(\vec\gamma^\alpha).
\ee
From comparing \eq{soltau}, \eq{soly+}, \eq{susoly-1}
to the minimum values in \eq{mintau}  and 
using the definition of the charge bilinears in \eq{tdualinv} and 
\eq{sdualcomb} it is seen that we obtain the following equations
\bea{mincorresp}
\gamma^1|_{*} = \tau|_{*}, \qquad
\gamma^2|_{*} = y^+|_{*}, \qquad
 \gamma^3|_{*} = -\frac{1}{y^-}|_{*}.
\eea
From \eq{stattot} and \eq{valminsent} we see that to ${\cal O}(1/Q^2)$,
the total statistical entropy is given by
\bea{totmin}
-\tilde\Gamma_B(\vec\gamma^1, \vec\gamma^2, \vec\gamma^3)|_*
&=&  
\pi \sqrt{Q^2P^2 - (Q\cdot P)^2 } \\ \nonumber
 & & 
-\sum_{\alpha =1}^3\left. \ln\left[ \vartheta_2(\gamma^\alpha)^4 
\vartheta_2(-\bar\gamma^\alpha )^4
(2\gamma_2^\alpha)^2 \right]\right|_* + \hbox{constant} + {\cal O}(1/Q^2).
\eea
Using \eq{mincorresp} and \eq{totmin} we see that the statistical
entropy coincides with the entropy of the black hole to the next leading 
order. 

\section{Approximate partition function for the FHSV model}

The subleading corrections to the entropy from the the coefficient of the
Gauss-Bonnet in the FHSV model depends on the attractor values of the
T-moduli through the Bocherds-Enriques form \eq{fshvf1}. 
The complete dyon partition function should capture this dependence on 
the T-moduli. Here we will focus on the scaling limit \eq{scalimit} 
in which the electric charges are much larger than the magnetic charges 
and write down an approximate dyon partition function which captures 
the degeneracy in this limit.
\be{prtfn1}
d(Q, P) = \frac{K'}{2} \exp{i\pi(Q\cdot P)}\int_C 
d\tilde\rho d\tilde\sigma d\tilde v 
\frac{1}{\Phit_4(\tilde\rho, \tilde\sigma, \tilde v)} \exp\left[
-i\pi \left( \tilde\sigma Q^2 + \tilde\rho P^2 + 2 \tilde v Q\cdot P\right) 
\right],
\ee
where $Q^2\equiv Q\cdot Q$, $P^2 \equiv P\cdot P$, 
$\Phit_4$ is a function to be
specified, and $C$ is a three real dimensional subspace of the three complex
dimensional space labeled by $(\tilde\rho, \tilde\sigma, \tilde v)$ given by
\bea{domains}
{\rm Im}\, \tilde \rho = M_1 \quad  {\rm Im}\,  \tilde\sigma = M_2,\quad 
\quad {\rm Im}\, \tilde v = M_3, 
\cr
0 \leq {\rm Re}\, \tilde\rho \leq 1, 
\quad  0 \leq{\rm Re}\, \tilde\sigma \leq 2, \quad
0\leq{\rm Re}\, \tilde v \leq 1.
\eea
$M_1, M_2, M_3$ being fixed large positive numbers. The normalization 
constant in \eq{prtfn1} is given by $K' = 2^{-6}$. 
The $Sp(2, \ZZZ)$ modular form  $\Phi_4$ which occurs 
in \eq{prtfn1} is  by
\be{reltphi4}
\Phit_4 (\tilde\rho, \tilde\sigma, \tilde v) = \left(
\Phit_6(\tilde\rho, \tilde\sigma, \tilde v) 
\Phit_2(\tilde\rho, \tilde\sigma, \tilde v) \right)^{\frac{1}{2}},
\ee
where $\Phit_6$ is the modular form of 
weight $6$ which captures the degeneracy
of dyons for the $N=2$ CHL orbifold 
discussed in \cite{Jatkar:2005bh,David:2006ji}
Similarly  $\Phit_2$ is the modular form
of $Sp(2, \ZZZ)$ of weight $2$ which capture degeneracy of 
dyons for the $\ZZZ_2$ orbifold of type II theory 
constructed in \cite{David:2006ru}. 
From the definition of $\Phit_4$ in \eq{reltphi4} it is easily seen that
it is a modular form of weight $4$ under the subgroup
$\tilde G$ of $Sp(2, \ZZZ)$ defined in \cite{Jatkar:2005bh}. 
Thus we have
\be{modtranp}
\Phit_4\left( A\Omega + B)(C\Omega + D)^{-1} \right) = \hbox{det}\, 
(C\Omega +D)^4 \Phit_4(\Omega), \qquad
\left(
\begin{array}{cc}
A & B \\
C & D
\end{array}
\right) \in \tilde G.
\ee
We now list some properties of $\Phit_4$ which are discussed in the 
the appendix in detail.

\vspace{.5cm}
\noindent
{\emph{Properties of ${\Phit}_4$}}
\vspace{.5cm}

\begin{enumerate}
\item
$\Phit_4(\tilde\rho, \tilde\sigma, \tilde v)$ is an analytic function 
in $\tilde\rho, \tilde\sigma, \tilde v$ with second order zeros at
\bea{loczero}
\left( n_2(\tilde \sigma\tilde\rho -\tilde v^2) +
 b\tilde v + n_1 \tilde\sigma -
\tilde\rho m_1 + m_2\right) =0, \\ \nonumber
{\rm for}\; m_1\in 2\ZZZ, m_2, n_2\in\ZZZ, b\in 2\ZZZ+1, 
m_1n_1 +m_2n_2 + \frac{b^2}{4} = \frac{1}{4}. 
\eea
It has simple poles at
\bea{locpol}
& &\left( n_2(\tilde \sigma\tilde\rho -\tilde v^2) +
 b\tilde v + n_1 \tilde\sigma -
\tilde\rho m_1 + m_2\right) =0, \\ \nonumber
& &{\rm for}\, m_1 \in 2\ZZZ+1, m_2, n_2 \in \ZZZ, n_1\in \ZZZ, b\in 2\ZZZ+1, 
m_1n_1 +m_2n_2 + \frac{b^2}{4} = \frac{1}{4}.
\eea
\item
$\Phit_4$ is invariant under the following $Sp(2,\ZZZ)$ transformations
\bea{sp2ztf}
\Phit_4(\tilde\rho', \tilde\sigma', \tilde v') = 
\Phit_4(\tilde\rho,\tilde\sigma,\tilde v), 
\\ \label{ttt1}
\hbox{for}\; \tilde\rho'= \tilde\rho + 4\tilde\sigma + 4 \tilde v -2, \quad
\tilde\sigma' = \tilde\sigma, \quad \tilde v' = 2\tilde\sigma + \tilde v, \\ 
 \label{ttt2}
\hbox{and} \; \tilde\rho'= \tilde\rho, \quad
\tilde\sigma' = 4\tilde\rho +\tilde\sigma+ 4\tilde v -2 , \quad 
\tilde v' = 2\tilde\sigma + \tilde v.
\eea
This property is due to the fact that $\Phit_4$ is invariant under the 
subgroup of $\tilde H$ defined in \eq{subgroup}. The transformation 
\eq{ttt1} corresponds to the choice $a= -1, b=2, c=0, d=-1$ in \eq{sghtrans}
and  \eq{ttt2} corresponds to the choice $a =1, b =0, c=-2, d=1$ in
\eq{sghtrans}.  The above $Sp(2, \ZZZ)$ transformations essentially 
correspond to generators of $\Gamma(2)$, but $\Phit_4$ is invariant 
under the subgroup \eq{subgroup} 
which is $\Gamma_0(2)$, this contains the group
$\Gamma(2)$. 
\item
 The expansion of $1/\Phit_4$ in terms of Fourier coefficients  given by
\be{fexpf}
\frac{K'}{\Phit_4(\tilde\rho,\tilde\sigma, \tilde v)} = 
\sum_{\stackrel{m, n, p}{m\geq -1, n \geq -1/2}}
e^{2\pi i( m \tilde\rho + n\tilde\sigma + p\tilde v)}
g(m , n, p),
\ee
with $g(m, n, p)$ being integers. 
$m , p\in \ZZZ$ while $n$ runs over integer multiples
of $1/2$
\item
To determine the asymptotic properties of the partition
function given in \eq{prtfn1} for large
charges we need the behavior of $\Phit_4$ near 
$\tilde\sigma\tilde\rho -\tilde v^2 + \tilde v =0$.
Near this surface $\Phit_4$ factorizes as
\be{facprop3}
\Phit_4(\tilde\rho, \tilde\sigma, \tilde v) 
= 4\pi^2 ( 2 v-\rho - \sigma)^2 v^2 f(\rho) f(\sigma) + {\cal O}(v^4),
\ee
where
\be{deffrho}
f^{(4)} (\rho) = \eta(2\rho)^{12},
\ee
and the relationship between the variables $\rho, \sigma, v$ 
and $\tilde\rho, \tilde\sigma, \tilde v $ is given by
the $Sp(2, \ZZZ)$ transformation \eq{chgvar} and \eq{chgvar2}.
\end{enumerate}

Just as in the case of $\Phit_0$ in the previous section, following the 
same logic and using the fact that $\Phit_4$ is also invariant under 
the $Sp(2, \ZZZ)$ transformations given in \eq{ttt1}and \eq{ttt2}, we see that
the integrand in \eq{prtfn1} is invariant under $\Gamma(2)_S$ symmetry.
This is what is expected of a partition function which aims to capture the 
dependence of the axion-dilaton moduli dependence of the subleading terms
in the entropy function. In fact from property since $\Phit_4$ is invariant
under the subgroup \eq{subgroup}, the partition function \eq{prtfn1} is
invariant under the larger group $\Gamma_0(2)$. 

We shall now compute the statistical entropy from the
approximate partition function for the FHSV model given in \eq{prtfn1}. 
We will see that the statistical entropy from this partition function 
captures the axion-dilaton moduli dependence of the entropy
\eq{fshvf1}. To obtain the statistical entropy we will follow the 
method of \cite{David:2006yn}.
of the dyons in the FHSV model. The value of this function at its
extremum gives the statistical entropy, the logarithm of the degeneracy
of states corresponding to the given set of charges. 
We see that the black hole entropy function evaluated in section 4
agrees precisely with the statistical entropy.
Following \cite{David:2006yn,Dijkgraaf:1996it,LopesCardoso:2006bg}
one can show that the dominant contribution to the above integral comes
form the pole at
\be{poleq}
\tilde\sigma\tilde\rho - \tilde v^2 + \tilde v =0.
\ee
The behavior of $\Phit_4$ near this zero is given by \eq{facprop3}, which 
is identical to the corresponding relation $(4.17)$ with $k\rightarrow 4$ and
$f^{(k)}(\rho) \rightarrow f(\rho)$. 
Thus following an analysis identical 
to that in \cite{David:2006yn} we can conclude that for large 
charges the statistical entropy
$S_{\rm{stat}} (Q, P)$ defined as the logarithm of the degeneracy $d(Q, P)$
is obtained by minimizing the statistical entropy function
\bea{statentrop}
-\tilde\Gamma_B(\vec\tau) &=& \frac{\pi}{2\tau_2}| Q + \tau P|^2 
-\ln (f^{(4)} (\tau)) - \ln( f^{(4)}(-\bar\tau)) - 6 \ln (2\tau_2) + 
{\hbox{constant}} \nonumber \\ 
& &  + 
{\cal O}(1/Q^2, 1/P^2),
\eea
where 
\be{deffnct}
f^{(4)}(\tau) = \eta(2\tau)^{12}. 
\ee
We can then obtain the statistical entropy by first minimizing the the function
$-\tilde\Gamma_B$ with respect to the real and 
imaginary parts of $\tau$ and then 
evaluating the value of the statistical entropy function at this 
critical point.
This gives
\be{fshventm}
-\tilde\Gamma_B(\vec\tau)|_* = \pi \sqrt{Q^2P^2 - (Q\cdot P)^2} 
- \ln\left[ f^{(4)}(\tau)f^{(4)}(-\bar\tau) (2\tau_2)^6 \right]|_*, 
\ee
where 
\be{crtvtau}
\tau_1|_* = -\frac{Q\cdot P}{P^2}, \qquad
\tau_2|_* = \frac{\sqrt{Q^2P^2 - (Q\cdot P)^2}}{P^2}.
\ee
Comparing \eq{fshvf1} and \eq{fshventm} using \eq{soltauf} and 
\eq{crtvtau} we see that the approximate partition function given in 
\eq{prtfn1} captures the contribution of the axion-dilaton moduli in
the subleading terms of the black hole entropy obtained from the
Gauss-Bonnet term. Therefore in the scaling limit given in \eq{scalimit} 
where the T-moduli contribution to the entropy is subleading, the approximate
partition function given in \eq{prtfn1} agrees with 
the black hole entropy \eq{scalentfn} ignoring ${\cal O}(Q^0, P^0)$ 
terms.

We note that the vector multiplet moduli space of the FHSV model
factorizes into the axion-dilaton dependence and the T-moduli dependence.
We also have seen that  the partition function in \eq{prtfn1} captures the 
subleading contribution to the dyon entropy form the axion-dilaton dependence
of the Gauss-Bonnet term.
Thus we can conclude that a product of $\Phit_4$ with a suitable 
function should capture the degeneracy of dyons in the FHSV model.
It will be interesting to determine this function using clues from the 
corrections to the entropy function given in \eq{fshvf1} and the 
attractor values of the T-moduli given in \eq{solyiy-} and \eq{solyfy+}.

\acknowledgments

The author wishes to thank N. Banerjee, A. Dabholkar, R. Gopakumar,
 B. Pioline,  B. Sahoo for useful comments and 
D. Jatkar,   A. Sen for extensive discussions and for a careful reading
of the manuscript.  
He thanks A. Dabholkar and LPTHE,
Jussieu  for hospitality  during an extended visit  where part of this 
work was done. Support from the XIth  plan of the DAE 
(Project No: 11-R\&D-HRI-5.02-0304) is acknowledged.

\appendix
\section{'t Hooft symbols for $SO(2,2)$}

We obtain the `t Hooft symbols for $SO(2,2)$ from the corresponding 
symbols for $SO(4)$ given by \cite{Hooft:1976fv} by the prescription:
$\eta_{a\mu\nu} 
\rightarrow (i)^{\delta_{a1} - \delta_{a2} + \delta_{\mu 3} +\delta_{\nu 4}}
\eta_{a\mu\nu}$ with an identical prescription for the 
anti-self dual `t Hooft symbols. 
This prescription takes care of the appropriate signature changes 
required in going from $SO(4)$ to $SO(2,2)$. 
Using this prescription we obtain 
the following self dual 't Hooft symbols for $SO(2,2)$, $\eta_{a;ij}$ with 
$a= 1, 2, 3$ and $i, j = 1, \cdots 4$  are defined as
\bea{defseld}
\eta_{1;23} = -1, &\qquad \eta_{2;31} = 1, &\qquad \eta_{3;12} =1, \\ \nonumber
\eta_{1;32} = 1, &\qquad \eta_{2;13} =-1, &\qquad \eta_{3;21} = -1, \\ \nonumber
\eta_{1;41} = 1, &\qquad \eta_{2;42} =-1, &\qquad \eta_{3;43} = +1, \\ \nonumber
\eta_{1;14} =-1, &\qquad \eta_{2;24} = 1, &\qquad \eta_{3;34} = -1.
\eea
All the remaining components vanish.
The anti-self dual 't Hooft symbols for $SO(2,2,)$ $\tilde\eta_{a;ij}$ are 
defined as 
\bea{adefseld}
\tilde\eta_{1;23} = -1, &\qquad \tilde\eta_{2;31} = 1, &\qquad 
\tilde\eta_{3;12} =1, \\ \nonumber
\tilde\eta_{1;32} = 1, &\qquad \tilde\eta_{2;13} =-1, &\qquad 
\tilde\eta_{3;21} = -1, \\ \nonumber
\tilde\eta_{1;41} = -1, &\qquad \tilde\eta_{2;42} =1, &\qquad 
\tilde\eta_{3;43} = -1, \\ \nonumber
\tilde\eta_{1;14} =1, &\qquad \tilde\eta_{2;24} = -1, &\qquad \eta_{3;34} = 1.
\eea
They satisfy the following identities:
\bea{thofid}
\eta_{a;ij} &=& \frac{1}{2}\epsilon_{ijkl} \eta_a^{\,kl}, \qquad
\tilde\eta_{a;ij} = -\frac{1}{2} \epsilon_{ijkl} \tilde\eta_a^{\, kl}, \\
\eta_{a;ij} &=& -\eta_{aji}, \qquad \tilde\eta_{a;ij} = -\tilde\eta_{a;ji}, \\
\eta_{a;ij} \eta_{b}^{\,ij} &=& 4 n_{ab}, \qquad 
\tilde\eta_{a;ij} \tilde\eta_{b}^{\,ij} = 4 n_{ab},\\
\eta_{a;ij} \eta^a_{\,kl} &=& 
L_{ik}L_{jl} - L_{il}L_{jk} + \epsilon_{ijkl}, \label{a6}\\
\tilde\eta_{a;ij} \tilde\eta^a_{\,kl} &=& 
L_{ik}L_{jl} - L_{il}L_{jk} - \epsilon_{ijkl}.
\eea
Raising and lowering of $i, j$ indices are performed using the 
metric $L_{ij} = {\rm {Dia}} ( 1, 1, -1,-1)$. The $S0(2,1)$ metric 
$n_{ab}$ is given by $n_{ab} = {\rm{Dia}}( -1, -1, 1)$ and the raising and 
lowering of $a, b$ indices are performed by the metric $n_{ab}$. 
We define the following combination of the 't Hooft symbols as
\be{comthoft}
\eta_{\pm ij} = \pm \eta_{2;ij} + \eta_{3;ij}, \qquad
\tilde\eta_{\pm ij} = \pm \tilde\eta_{2;ij} + \tilde\eta_{3;ij}.
\ee
\section{Attractor values for the T-moduli in  the FHSV model}

In this part of the appendix we provide the details of the 
calculations which leads to the attractor values of the 
T-moduli in the FHSV model given in \eq{solyiy-} and 
\eq{solyfy+}.
We start with the entropy function of the FHSV model given in 
\eq{compent}, to determine the 
values of the T-moduli at the attractor point it is 
sufficient to focus on the first two terms of \eq{compent}. 
The first two terms are identical except for the exchange of $\tau
\leftrightarrow \bar\tau$ in the numerator. 
Our strategy for minimizing with
respect to the T-moduli is similar to the one we followed in the case of the
STU model. We will just focus on the first term of \eq{compent}, this is 
is given by
\be{enfshtmod}
F_{\rm{ T} } = \frac{\pi}{2} \frac{|(Q + \tau P)\cdot w|^2}{\tau_2 Y},
\ee
and minimize this term with respect to the T-moduli. We will see that the 
attractor values of the moduli are independent of the axion-dilator moduli
$\tau$.
Therefore these values of the attractor moduli minimize the second term in 
\eq{compent} simultaneously. This is because the second term is the same as
the first term with $\tau\rightarrow \bar\tau$. 
For convenience we introduce the variables
\bea{convarf}
& &n_1 = (Q^{12} + Q^{10}) + \tau( P^{12} + P^{10}) , \quad
n_2 = (Q^{11} +  Q^{9}) + \tau( P^{11} + P^{9}) \\ \nonumber
& &w_1' =( Q^{12} -  Q^{10} )+ \tau( P^{12} - P^{10}) , \quad
w_2' = (Q^{9} - Q^{11} )+ \tau( P^{9} - P^{11}), \\ \nonumber
& &q^i = Q^i + \tau P^i, \quad i = 1, \cdots 8.
\eea
With these variables one can write \eq{enfshtmod} as 
\bea{rwriteenf}
F_{\rm{ T} } &=& -4 \frac{|E|^2}{ D} \\ \nonumber
\hbox{where}\; 
E &=& - \vec q\cdot \vec y - w_1'\frac{\vec y^2}{4} - n_1 
+ \frac{y^+y^-}{2} w_1' + \frac{y^+}{\sqrt{2}} w_2'
+ \frac{y^-}{\sqrt{2}} n_2, \\ \nonumber
\hbox{and}\;
D&=& 2 (y^+- \bar y^+) ( y^--\bar y^-) - 
\sum_{i=1}^8 (y^i-\bar y^i)^2, 
\eea
where we have substituted for $w$ in terms of $y$ using \eq{relwy} and used the
variables given in \eq{convarf}. 
Minimizing with respect to $y^+$ we get the equation
\be{miny+eq}
\frac{y^+w_1'}{2} + \frac{n_2}{\sqrt{2}} = 2\frac{E}{D} ( y^+ -\bar y^+). 
\ee
The above equation can be simplified by 
further substituting for $E$ and $D$ in 
terms of the $y$'s, this results in 
\bea{miny+eq1}
\bar y^+& =& \frac{n_1 - \frac{y^-n_2}{\sqrt{2}} }
{ \frac{y^-w_1'}{2} + \frac{w_2'}{\sqrt{2}}
} 
  - \sum_{i=1}^8 \frac{ (y^i-\bar y^i)^2}{2 (y-\bar y)} 
+ \frac{ \vec q\cdot \vec y + \frac{\vec y^2}{4} w_1' }
{\frac{y^-w_1'}{2} + \frac{w_2'}{\sqrt{2}} }. 
\eea
Minimizing with respect to $y^-$ and $y^i$ we obtain the equations
\bea{miny-eq}
\frac{y^-w_1'}{2} + \frac{w_2'}{\sqrt{2}} &=& 2\frac{E}{D} ( y^- -\bar y^-), \\
\label{minyieq}
q^i + \frac{w_1'}{2} y^i &= &2\frac{E}{D} ( y^i -\bar y^i).
\eea
From \eq{miny+eq} , \eq{miny-eq} and \eq{minyieq}
 we can solve for $y^+$, $y^-$ and $y^-$
in terms of the ratio
\be{defratio}
R = 2\frac{E}{D}.
\ee
These are given by
\bea{solyinR}
y^+ &=& \frac{1}{ (\frac{w_1'}{2} - R) (\frac{\bar w_1'}{2} - \bar R) - \bar R R}
\left( - \frac{n_2}{ 2\sqrt{2}} \bar w_1' + \frac{n_2}{\sqrt{2}} \bar R + 
\frac{\bar n_2 }{\sqrt{2}} R \right), \\ \nonumber
y^- &=&\frac{1}{ (\frac{w_1'}{2} - R) (\frac{\bar w_1'}{2} - \bar R) - \bar R R}
\left( - \frac{w_2'}{ 2\sqrt{2}} \bar w_1' + \frac{w_2'}{\sqrt{2}} \bar R + 
\frac{\bar w_2' }{\sqrt{2}} R \right) \\ \nonumber
y^i &=& \frac{1}{ (\frac{w_1'}{2} - R) (\frac{\bar w_1'}{2} - \bar R) - \bar R R}
\left( - \frac{ q^i}{2} \bar w_1' + q^i \bar R + \bar q^i R\right).
\eea
From the above equations it is possible to evaluate the following ratios
which 
are independent of $R$. 
\bea{indratio}
\frac{y^-- \bar y^-}{y^+ -\bar y^+} 
 &=& \frac{ \bar w_2' w_1' - w_2'\bar w_1'}{ \bar n_2 w_1' - n_2\bar w_1'}, \\ 
 \label{indratio2}
 \frac{ y^i - \bar y^i}{y^+ -\bar y^+}
 &=& \sqrt{2} \frac{ \bar q^i w_1' - q^i \bar w_1'}{ \bar n_2 w_1' - n_2 \bar w_1'}.
\eea
It is easy to see that the ratios on the 
right hand side of the above equations 
are independent of the axion-dilaton moduli $\tau$ on substituting the 
definitions \eq{convarf}. 
Now from \eq{miny+eq} and \eq{miny-eq} and \eq{indratio}
we can obtain the value of $y^-$ in terms of $y^+$
this is given by
\bea{soly-a}
y^- &=& y^+ \frac{\bar w_2w_1' -w_2\bar w_1'}{\bar n_2 w_1' - n_2\bar w_1'}
+ \sqrt{2}\frac{n_2 \bar w_2' - w_2'\bar n_2}{\bar n_2 w_1' - n_2 \bar w_1'},
\\ \nonumber
&=&  y^+ \frac{W_2\tilde W_1 - \tilde W_2  W_1}
{N_2 \tilde W_1 - \tilde N_2 W_1} + \sqrt{2}
\frac{\tilde N_2 W_2 - N_2 \tilde W_2}{N_2 \tilde W_1 - \tilde N_2 W_1}.
\eea
where the $N$, $W$ and $\tilde N$ and $\tilde W$ are defined in 
\eq{combchg}. Similarly, from \eq{miny+eq} and \eq{minyieq} and \eq{indratio2}
we obtain $y^i$ in terms of $y^+$
\bea{solyia}
y^i & =& \sqrt{2} y^+ \frac{\bar q^i w_1' - q^i \bar w_1'}{ \bar n_2 w_1' -n_2 \bar w_1'}
+ 2 \frac{ n_2 \bar q^i -\bar n_2 q^i}{ \bar n_2 w_1' -n_2 \bar w_1'}, \\ \nonumber
&=& \sqrt{2} y^+
\frac{Q^i \tilde W_1 - P^i W_1}{ N_2 \tilde W_1 - \tilde N_2 W_1}
+ 2 \frac{\tilde N_1 Q^i - N_2P^i}{ N_2 \tilde W_1 - \tilde N_2 W_1}.
\eea
We can now substitute the solutions for $y^i$ and $y^-$ in terms of 
$y^+$ in the equation \eq{miny+eq1} and obtain the following equation 
for $y^+$ after some manipulations
\bea{ccy+}
& &\tilde A \bar y^+ y^+ + \tilde B (y^+ + \bar y^+) + \tilde C =0,
\\ \nonumber
& &\tilde A = - \frac{w_1}{2} \left (
\frac{\bar w_2' w_1' - w_2' \bar w_1'}{\bar n_2 w_1' - n_2 \bar w_1'}
- \sum_{i=1}^8 \frac{( \bar q^i  w_1' - q^i \bar w_1')^2 }{(\bar n_2 w_1' - n_2 \bar w_1')^2}
\right),
\\ \nonumber
& &\tilde B  = - \frac{n_2}{\sqrt{2}} 
\left (
\frac{\bar w_2' w_1' - w_2' \bar w_1'}{\bar n_2 w_1' - n_2 \bar w_1'}
- \sum_{i=1}^8 \frac{( \bar q^i  w_1' - q^i \bar w_1')^2 }{(\bar n_2 w_1' - n_2 \bar w_1')^2}
\right), \\ \nonumber
& &\tilde C = 
\frac{n_1( \bar n_2 w_1' - n_2 \bar w_1') - n_2 (n_2\bar w_2' - w_2' \bar n_2)}{
\bar n_2 w_1' - n_2 \bar w_1'}
+ n_2 \frac{ \sum_{i=1}^8 (n_2 \bar q^i -\bar n_1 q^i)( w_1' \bar q^i - q^i \bar w_1')}{(
\bar n_2 w_1' - n_2 \bar w_1')^2}.
\eea
Using the above equation and its complex conjugate one can obtain a 
relationship between $y^+$ and $\bar y^+$ by getting rid of the quadratic 
term $\bar y^+ y^+$ in \eq{ccy+}. This is given by
\bea{y+bary+}
& &\frac{\bar y^+}{\sqrt{2}} 
= \frac{y^+}{\sqrt{2}}  +\\ \nonumber
\; & &  \frac{
(\bar n_2 w_1' - n_2 \bar w_1')(\bar w_1' n_1 -w_1'\bar n_1 + n_2\bar w_2'
-w_2'\bar n_2) - \sum_{i=1}^8
(n_2 \bar q^i -\bar n_1 q^i)( w_1' \bar q^i - q^i \bar w_1')}{
(\bar n_2 w_1' - n_2 \bar w_1')( w_1\bar w_2' -w_2'\bar w_1') 
- \sum_{i=1}^8 ( \bar q^i  w_1' - q^i \bar w_1')^2 }.
\eea
Substituting this relationship in \eq{ccy+} and writing the equation
in terms of only $y^+$ we obtain the quadratic equation
\bea{solyfy+a}
& &A\left( \frac{y^+}{\sqrt{2}}\right)^2 + B \frac{y^+}{\sqrt{2}} 
+ C = 0, \\ \nonumber
& &A = \tilde W_2 W_1 - W_2\tilde W_1 + \sum_{i=1}^8 
\frac{(Q_i \tilde W_1 - P_i W_1)^2 }{ N_2 \tilde W_1 - \tilde N_2 W_1}, \\
\nonumber
& & B = (\tilde W_2 N_2 - W_2 \tilde N_2) -( \tilde N_1 W_1 - N_1 \tilde W_1)
+ \sum_{i=1}^8
\frac{( \tilde N_2 Q_i -P_i N_2)(\tilde W_1 Q_i - P_i W_1) }{
 N_2 \tilde W_1 - \tilde N_2 W_1},
\\ \nonumber
& &C= N_1 \tilde N_2  - \tilde N_1 N_2.
\eea
Here we have used the definitions \eq{convarf} and \eq{combchg}.
Note that from \eq{soly-a}, \eq{solyia} and 
\eq{solyfy+a} the solution fo the T-moduli $y^+, y^-
 y^i$  are independent of the axion-dilaton moduli $\tau$ and thus solve the 
attractor equations for the entropy function \eq{compent}.  
Note that the above procedure of solving the attractor equations 
for the T-moduli generalize trivially to any model with 
vector multiplet moduli space 
\be{genvemod}
{\cal M}_V = \frac{SU(1,1)}{U(1) } \times \frac{SO(2, n)}{SO(2)\times SO(n)}
\ee

\section{Properties of Siegel modular forms}

In this section we detail the construction and properties  of the
Siegel modular forms $\Phit_0$ and $\Phit_4$ which are used to
 write down the dyon partition functions for the two ${\cal N}=2$
models discussed in this paper.
From and \eq{defphi0}, \eq{reltphi4} we see that these  modular
forms are defined in terms of the modular forms $\Phit_2$, $\Phi_6$
and $\Phit_2'$. We first recall the 
construction and properties of $\Phit_2$ and $\Phit_6$ which was constructed 
to capture the degeneracy of 
dyons in a $\ZZZ_2$ orbifold of type II theory and the
$\ZZZ_2$ CHL orbifold respectively in \cite{David:2006ru}, 
\cite{David:2006ji}. 

\vspace{.5cm}
\noindent
{\underline{$\Phit_6(\tilde\rho,\tilde\sigma,\tilde v)$}}
\vspace{.5cm}

\noindent
The infinite product representation of 
$\Phi_6$ is given by \cite{David:2006ji}
\bea{defphi6}
\Phit_6 (\tilde \rho ,\tilde \sigma, \tilde v) &=& -
\frac{1}{16} e^{(2\pi i( \frac{\tilde\sigma}{2} +\tilde
\rho + \tilde v) )} \times  \\ \nonumber
&\,&\prod_{r =0}^1 \prod_{ 
\stackrel{k'\in \zzz + \frac{r}{2}, l, j \in \zzz} 
{k', l \geq 0 , j<0 {\rm for}\, k'=l=0} }
\left( 1 - \exp(2\pi i ( k' \tilde\sigma + l\tilde\rho + j \tilde v) 
\right)^ { \sum_{s=0}^1 (-1)^{sl} c^{(r, s)}_6 ( 4k'l -j^2)} ,
\eea
where the coefficients $c^{(r,s)}_6$ are defined by the expansion
\be{defcrs6}
F^{(r,s)}_6 (\tau, z) = \sum_{b\in\zzz, n} c^{(r, s)}_6(4n -b^2) q^n e^{2\pi i b},
\ee
here $n \in \ZZZ$ for $r=0$ and $\frac{1}{2}\ZZZ$ for $r=1$. The expressions
for various values of $(r, s)$ are as follows:
Let 
\be{expfrs6}
F^{(r,s)}_6(\tau, z) = 
h_{6;0}^{(r,s)}(\tau) \vartheta_3(2\tau, 2z) + h_{6;1}^{(r,s)}(\tau)
\vartheta_2(2\tau, 2z),
\ee
here we list these functions
\bea{defhlrs6}
h_{6;0}^{(0,0)}(\tau) &=& 
8 \frac{\vartheta_3(2\tau, 0)^3}{\vartheta_3(\tau, 0)^2\vartheta_4(\tau, 0)^2} 
+ 2\frac{1}{\vartheta_3(2\tau, 0)}, \cr
h_{6;1}^{(0,0)}(\tau) &=& -8\frac{\vartheta_2(2\tau, 0)^3}
{\vartheta_3(\tau, 0)^2\vartheta_4(\tau, 0)^2} + 
2\frac{1}{\vartheta_2(2\tau, 0)}, \cr
h_{6;0}^{(0,1)}(\tau) &=& 2 \frac{1}{\vartheta_3(2\tau, 0)}, \qquad
h_{6;1}^{(0,1)}(\tau) = 2 \frac{1}{\vartheta_2(2\tau, 0)}, \cr
h_{6;0}^{(1,0)}(\tau) &=& 
4 \frac{\vartheta_3(2\tau, 0)}{\vartheta_4(\tau, 0)^2}, 
\qquad h_{6;1}^{(1,0)}(\tau) = 
-4\frac{\vartheta_2(2\tau, 0)}{\vartheta_4(\tau, 0)^2},
\cr
h_{6;0}^{(1,1)}(\tau)
&=& 4 \frac{\vartheta_3(2\tau, 0)}{\vartheta_3(\tau, 0)^2}, 
\qquad h_{6;1}^{(1,0)}(\tau) = 
4\frac{\vartheta_2(2\tau, 0)}{\vartheta_3(\tau, 0)^2}.
\eea
We can now define the coefficients $c^{(r,s)}(u)$ through the expansions
\be{exphlrs6}
h_{6;0}^{(r,s)}(\tau) = \sum_n c^{(r,s)}_6(4n)q^n, \qquad
h_{6;1}^{(r,s)}(\tau) = \sum_n c^{(r,s)}_6(4n) q^n.
\ee
From \eq{defhlrs6} we see that in the expansion 
of $h_{6;l}^{(r,s)}$, 
$n \in \ZZZ - \frac{l}{4}$ for $r=0$ and 
$n \in \frac{1}{2}\ZZZ - \frac{l}{4}$ for $r=1$. 
Thus for given $(r,s)$ the $c^{(r,s)}(u)$ defined through the two
equations in \eq{exphlrs6} have non-overlapping set of arguments. 
Substituting \eq{exphlrs6} in \eq{expfrs6} we get \eq{defcrs6}
The properties of $\Phit_6$ can be studied by obtaining it as 
a result of a threshold like integral \cite{David:2006ji}. 
Here we review some of its important properties:
\begin{enumerate}
\item
$\Phit_6$ is a modular form of weight $6$ under the subgroup
$\tilde G$ of $Sp(2, \ZZZ)$ defined in \cite{Jatkar:2005bh}. Therefore
\bea{modtran6}
& &\Phit_6( (A\Omega + B) (C\Omega + D)^{-1} ) = \det( C\Omega + D)^6
\tilde\Phi_6(\Omega), \quad\left(
\begin{array}{cc}
A & B\\ C& D \end{array} \right) \in \tilde G. \nonumber \\
& &\Omega = \left(
\begin{array}{cc}
\tilde\rho  & \tilde v \\
\tilde v & \tilde \rho
\end{array} \right)
\eea
$\tilde G$ contains the subgroup $\tilde H$ \cite{Jatkar:2005bh} whose
elements are of the form:
\be{subgroup}
\left(\begin{array}{cccc}
a & -b& b & 0 \\
-c & d & 0& c \\
0 & 0 & d & c \\
0 & 0 & b & a 
\end{array} \right) , \quad ad - bc =1, c \in 2\ZZZ.
\ee 
From \eq{modtran6}  we see that 
$\Phit_6$ is invariant under the $Sp(2, \ZZZ)$ transformation 
given in \eq{subgroup}.  This gives that for 
\bea{sghtrans}
\tilde \rho' &=& a^2\tilde\rho + b^2 \tilde\sigma - 2 ab \tilde v + ab, \\ \nonumber
 \tilde\sigma' &=& c^2 \tilde\rho + d^2 \tilde\sigma - 2 cd\tilde v + cd, \\
 \nonumber
\tilde v' &=& -ac \tilde\rho - bc \tilde\sigma + (ad + bc) \tilde v - bc,
\eea
we have 
\be{invphi6mod}
\Phit_6(\rho', \sigma' v') = \Phi_6(\rho, \sigma, v) \quad {\hbox{for}} \quad
\left( \begin{array}{cc}
a & b \\
c & d \end{array}\right),  \quad ad -bc =1, c\in 2\ZZZ.
\ee
Thus these group of matrices belong to $\Gamma_0(2)$. 
\item
From examining the coefficients $c^{(r,s)}_6$ defined by the expansions
in \eq{defhlrs6} it can be seen that
\be{evintp}
c^{(r,0)}_6( u) \pm c^{r,1}_6(u) \in 2\ZZZ
\ee
\item
$\Phit_6$ has second order zeros at 
\bea{loczero6}
\left( n_2(\tilde \sigma\tilde\rho -\tilde v^2) + b\tilde v + n_1 
\tilde\sigma -
\tilde\rho m_1 + m_2\right) =0, \\ \nonumber
{\rm for}\; m_1\in 2\ZZZ, m_2, n_2\in\ZZZ, b\in 2\ZZZ+1, 
m_1n_1 +m_2n_2 + \frac{b^2}{4} = \frac{1}{4} .
\eea
\item
In the limit $\tilde v \rightarrow 0$, $\Phit_6$ factorizes as
\bea{facprop6}
\Phit_6 (\tilde\rho, \tilde\sigma, \tilde v)_{\tilde v\rightarrow 0} =
-\frac{1}{4} \pi^2 \tilde v^2 \eta(\tilde\rho)^8 \eta(2\tilde\rho)^8
\eta(\frac{\tilde\sigma}{2})^8 
\eta(\tilde\sigma)^8.
\eea
\item
Under the $Sp(2,\ZZZ)$ transformation given in \eq{chgvar} $\tilde\Phi_6$ is 
related to the $Sp(2, \ZZZ)$ modular form of weight 6 by 
\bea{dualprop6}
\Phit_6(\tilde\rho, \tilde\sigma, \tilde v) &=&
\tilde\sigma^{-6} \Phi_6\left(\tilde\rho - \frac{\tilde v^2}{\tilde\sigma}, 
\frac{\tilde\rho\tilde\sigma - (\tilde v -1)^2}{\tilde\sigma}, 
\frac{\tilde\rho\tilde\sigma - \tilde v^2 + \tilde v}{\tilde \sigma}
\right), \\ \nonumber 
& =& \tilde \sigma^{-6}
\Phi_6(  \rho,\sigma, v), 
\eea
where $\Phi_6$ is defined by
\bea{defphio6}
\Phi_6(\rho, \sigma, v) &=& - \exp(2\pi i (\rho+ \sigma + v)) \\ \nonumber
&\,& \prod_{r,s=0}^1 \prod_{\stackrel{(k, l, b)\in \zzz}{k, l\geq 0, b<0\, 
\hbox{for}\, k=l=0}} 
\left\{ 1 - (-1)^r \exp( 2\pi i (k \sigma + l\rho + b v)) \right\}
^{ c^{(r,s)}_6 (4kl -b^2)},
\eea
$(\tilde\rho,\tilde\sigma,\tilde v)$ is related $(\rho, \sigma, v)$ by
\eq{chgvar}
$\Phi_6$ has the factorization property in the limit $v\rightarrow 0$
\be{fprop6}
\Phi_6(\rho, \sigma, v )_{v\rightarrow 0} = 4\pi^2 v^2 \eta(\sigma)^8 
\eta(2\sigma)^8 \eta(\rho)^8 \eta(2\rho)^8.
\ee
\item
Using \eq{dualprop6} and the factorization property 
\eq{fprop6} we see that when $\tilde\rho\sigma -\tilde v^2 + \tilde v \sim 0$
$\tilde \Phi_6$ factorizes as 
\be{fprops6}
\tilde\Phi_6(\tilde\rho, \tilde\sigma, \tilde v) \sim
4\pi^2 ( 2 v -\rho -\sigma)^2 v^2 \eta(\sigma)^8 
\eta(2\sigma)^8 \eta(\rho)^8 \eta(2\rho)^8.
\ee
\end{enumerate}

\vspace{.5cm}
\noindent
{\underline{$\Phit_2(\tilde\rho,\tilde\sigma, \tilde v)$}}
\vspace{.5cm}

\noindent
The infinite product representation of 
$\Phit_2$ is given by \cite{David:2006ru}
\bea{defphi2}
\Phit_2 (\tilde \rho ,\tilde \sigma, \tilde v) &=& -
\frac{1}{2^8} e^{(2\pi i(  \tilde
\rho + \tilde v) )} \times  \\ \nonumber
&\,&\prod_{r =0}^1 \prod_{ 
\stackrel{k'\in \zzz + \frac{r}{2}, l, j \in \zzz} 
{k', l \geq 0 , j<0 {\rm for}\, k'=l=0} }
\left( 1 - \exp(2\pi i ( k' \tilde\sigma + l\tilde\rho + j \tilde v) 
\right)^ { \sum_{s=0}^1 (-1)^{sl} c^{(r, s)}_2 ( 4k'l -j^2)}, 
\eea
where the coefficients $c^{(r,s)}_2$ are defined by the expansion
\be{defcrs2}
F^{(r,s)}_2 (\tau, z) = \sum_{b\in\zzz, n} c^{(r, s)}_2(4n -b^2) q^n e^{2\pi i b},
\ee
here $n \in \ZZZ$ for $r=0$ and $\frac{1}{2}\ZZZ$ for $r=1$. The expressions
for various values of $(r, s)$ are as follows:
Let 
\be{expfrs2}
F^{(r,s)}_2(\tau, z) = 
h_{2;0}^{(r,s)}(\tau) \vartheta_3(2\tau, 2z) + h_{2;1}^{(r,s)}(\tau),
\vartheta_2(2\tau, 2z).
\ee
here we list these functions
\bea{defhlrs2}
h_{2;0}^{(0,0)}(\tau) &=& 0 , \qquad
h_{2;1}^{(0,0)}(\tau) =0,  \\ \nonumber
h_{2;0}^{(0,1)}(\tau) &=& 4 \frac{1}{\vartheta_3(2\tau, 0)}, \qquad
h_{2;1}^{(0,1)}(\tau) = 4 \frac{1}{\vartheta_2(2\tau, 0)}, \cr
h_{2;0}^{(1,0)}(\tau) &=& 
8 \frac{\vartheta_3(2\tau, 0)}{\vartheta_4(\tau, 0)^2}, 
\qquad h_{2;1}^{(1,0)}(\tau) = 
-8\frac{\vartheta_2(2\tau, 0)}{\vartheta_4(\tau, 0)^2},
\cr
h_{2;0}^{(1,1)}(\tau)
&=& 8 \frac{\vartheta_3(2\tau, 0)}{\vartheta_3(\tau, 0)^2}, 
\qquad h_{6;1}^{(1,0)}(\tau) = 
8\frac{\vartheta_2(2\tau, 0)}{\vartheta_3(\tau, 0)^2}.
\eea
We can now define the coefficients $c^{(r,s)}(u)$ through the expansions
\be{exphlrs2}
h_{2;0}^{(r,s)}(\tau) = \sum_n c^{(r,s)}_2(4n)q^n, \qquad
h_{2;1}^{(r,s)}(\tau) = \sum_n c^{(r,s)}_2(4n) q^n.
\ee
From \eq{defhlrs2} we see that in the expansion 
of $h_{2;l}^{(r,s)}$, 
$n \in \ZZZ - \frac{l}{4}$ for $r=0$ and 
$n \in \frac{1}{2}\ZZZ - \frac{l}{4}$ for $r=1$. 
Thus for given $(r,s)$ the $c^{(r,s)}(u)$ defined through the two
equations in \eq{exphlrs2} have non-overlapping set of arguments. 
Substituting \eq{exphlrs2} in \eq{expfrs2} we get \eq{defcrs2}
The properties of $\Phit_2$ can be studied by obtaining it as 
a result of a threshold like integral \cite{David:2006ru}. 
Here we review some of its important properties:
\begin{enumerate}
\item
$\Phit_2$ is a modular form of weight $2$ under the subgroup
$\tilde G$ of $Sp(2, \ZZZ)$ defined in \cite{Jatkar:2005bh}. Therefore
\be{modtran2}
\Phit_2( (A\Omega + B) (C\Omega + D)^{-1} ) = \det( C\Omega + D)^2
\tilde\Phi_2(\Omega), \quad\left(
\begin{array}{cc}
A & B\\ C& D \end{array} \right) \in \tilde G.
\ee
Since the $\tilde G$ contains $\tilde H$ given in \eq{subgroup} using the
same arguments as in the case of $\Phit_6$ we see that
$\Phit_2$ is also invariant under the transformation 
\eq{sghtrans}. 
\item
From examining the coefficients $c^{(r,s)}_2$ defined by the expansions
in \eq{defhlrs2} it can be seen that
\be{evintp2}
c^{(r,0)}_2( u) \pm c^{r,1}_2(u) \in 2\ZZZ.
\ee
\item
$\Phit_2$ has second order zeros at 
\bea{loczero2}
& & \left( n_2(\tilde \sigma\tilde\rho -\tilde v^2) + b\tilde v + n_1 
\tilde\sigma -
\tilde\rho m_1 + m_2\right) =0, \\ \nonumber
& &{\rm for}\; m_1\in 2\ZZZ, m_2, n_2\in\ZZZ, n_1\in \ZZZ b\in 2\ZZZ+1, 
m_1n_1 +m_2n_2 + \frac{b^2}{4} = \frac{1}{4}. 
\eea
It also has a second order pole at 
\bea{locpol2}
&&\left( n_2(\tilde \sigma\tilde\rho -\tilde v^2) + 
b\tilde v + n_1 \tilde\sigma -
\tilde\rho m_1 + m_2\right) =0, \\ \nonumber
& &{\rm for}\, m_1 \in 2\ZZZ+1, m_2, n_2 \in \ZZZ, n_1\in \ZZZ, b\in 2\ZZZ+1, 
m_1n_1 +m_2n_2 + \frac{b^2}{4} = \frac{1}{4}.
\eea
\item
In the limit $\tilde v \rightarrow 0$, $\Phit_2$ factorizes as
\bea{facprop2}
\Phit_2 (\tilde\rho, \tilde\sigma, \tilde v)_{\tilde v\rightarrow 0} =
-\frac{\pi^2}{64} \tilde v^2 \frac{\eta(2\tilde\rho)^{16}}{\eta(\tilde\rho)^8}
\frac{\eta(\tilde\sigma/2)^{16}}{\eta(\tilde\sigma)^8}.
\eea
\item
Under the $Sp(2,\ZZZ)$ transformation given in \eq{chgvar} $\tilde\Phi_2$ is 
related to $\Phi_2$ $Sp(2, \ZZZ)$ modular form of weight 2 by 
\bea{dualprop2}
\Phit_2(\tilde\rho, \tilde\sigma, \tilde v) &=&
\tilde\sigma^{-2} \Phi_2\left(\tilde\rho - \frac{\tilde v^2}{\tilde\sigma}, 
\frac{\tilde\rho\tilde\sigma - (\tilde v -1)^2}{\tilde\sigma}, 
\frac{\tilde\rho\tilde\sigma - \tilde v^2 + \tilde v}{\tilde \sigma}
\right), \\ \nonumber 
& =& \tilde \sigma^{-2}
\Phi_2(  \rho,\sigma, v). 
\eea
where $\Phi_2$ is defined by
\bea{defphio2}
\Phi_2(\rho, \sigma, v) &=& - \exp(2\pi i (\rho+ \sigma + v)) \\ \nonumber
&\,& \prod_{r,s=0}^1 \prod_{\stackrel{(k, l, b)\in \zzz}{k, l\geq 0, b<0\, 
\hbox{for}\, k=l=0}} 
\left\{ 1 - (-1)^r \exp( 2\pi i (k \sigma + l\rho + b v)) \right\}
^{ c^{(r,s)}_2 (4kl -b^2)},
\eea
$(\tilde\rho,\tilde\sigma,\tilde v)$ is related $(\rho, \sigma, v)$ by
\eq{chgvar}
$\Phi_2$ has the factorization property in the limit $v\rightarrow 0$
\be{fprop2}
\Phi_2(\rho, \sigma, v )_{v\rightarrow 0} = 4\pi^2 v^2 
\frac{\eta(2\rho)^{16}}{\eta(\rho)^8} 
\frac{\eta(2\sigma)^{16}}{\eta(\sigma)^8}.
\ee
\item
Using \eq{dualprop2} and the factorization property 
\eq{fprop2} we see that when $\tilde\rho\sigma -\tilde v^2 + \tilde v \sim 0$
$\tilde \Phi_2$ factorizes as 
\be{fprops2}
\tilde\Phi_2(\tilde\rho, \tilde\sigma, \tilde v) \sim
4\pi^2 ( 2 v -\rho -\sigma)^2 v^2
\frac{\eta(2\rho)^{16}}{\eta(\rho)^8} 
\frac{\eta(2\sigma)^{16}}{\eta(\sigma)^8}.
\ee
\end{enumerate}

\vspace{.5cm}
\noindent
{\underline{$\Phit_2'(\tilde\rho,\tilde\sigma, \tilde v)$}}
\vspace{.5cm}

\noindent
$\Phit_2'$ is a $Sp(2,\ZZZ)$ modular form 
of weight 2 which is related to $\Phit_2$ by the 
following 
\be{re22'}
\Phit_2'(\tilde\rho,\tilde\sigma, \tilde v) =
\Phit_2(\frac{\tilde\sigma}{2}, 2\tilde\rho, \tilde v),
\ee
Though we can derive many of the properties of 
$\Phit_2'$ from the properties of $\Phit_2$ it is instructive to 
write down the threshold integral from which $\Phit_2'$ can be obtained
This threshold integral is given by
\be{rwrint}
{\cal I}(\tilde\rho, \tilde\sigma, \tilde v) = 
\sum_{l, r, s=0}^1 {\cal I}_{r,s,l},
\ee
where
\be{defirsl}
{\cal I}_{r, s,l} =
\int_{{\cal F}} \frac{d^2\tau}{\tau_2} 
\sum_{\stackrel{ m_2, n_2 \in \zzz, n_1\in \zzz/2}{m_1\in 2\zzz+ r, b \in 2\zzz+ l}}
q^{p_L^2/2} \bar q^{p_R^2/2}e^{(2\pi i n_1 s)} h_{2;l}^{(r,s)},
\ee
where ${\cal F}$ denotes the fundamental domain of $SL(2, \ZZZ)$ in the upper 
half plane, $ h_{2;l}^{(r,s)}$ are given in \eq{defhlrs2}.
Here
\bea{defqthre}
q &=& e^{2\pi i \tau}, \\
\frac{p_R^2}{2} &=& \frac{1}{4\det{\rm Im}\Omega} |- m_1 \tilde\rho +m_2 + 
n_1\tilde\sigma + n_2 (\tilde\sigma\tilde\rho -\tilde v^2) + b\tilde v|^2 , \\
\frac{p_L^2}{2} &=& \frac{1}{2} p_R^2 + m_1n_1 + m_2 n_2 + \frac{1}{4} b^2,
\\
\Omega &=& \left(
\begin{array}{cc}
\tilde \rho & \tilde v \\
\tilde v & \tilde\sigma
\end{array}
\right).
\eea
The integrals in \eq{rwrint} can be performed using the procedure
\cite{Dixon:1990pc,Kawai:1995hy,Harvey:1995fq,David:2006ji}.
The procedure involves evaluating the contribution of the zero orbit, the
degenerate orbit and the non-degenerate orbit of $SL(2, \ZZZ)$ separately.
The result is 
\bea{resulint}
{\cal I}
&=& -2 \ln\left[ 2^{16} \kappa( \hbox{det}\, \hbox{Im}\, \Omega)^2 \right]
- 2 \ln \Phit_2'(\tilde\rho, \tilde\sigma, \tilde v) -2 
\ln \bar{\Phit}_2'(\tilde\rho, \tilde\sigma, \tilde v),
\eea
where 
\be{defkappa}
\kappa = \left( \frac{8\pi}{3\sqrt{3}} e^{1-\gamma_E} \right)^4.
\ee
$\gamma_E$ is Euler's constant and $\Phit_2'$ is given by
\bea{defphi2'}
\Phit_2' (\tilde \rho ,\tilde \sigma, \tilde v) &=& -
\frac{1}{2^8} e^{(2\pi i(  \tilde
\frac{\sigma}{2} + \tilde v) )} \times  \\ \nonumber
&\,&\prod_{r =0}^1 \prod_{ 
\stackrel{k'\in 2\zzz + r, l\in \zzz/2, j \in \zzz} 
{k', l \geq 0 , j<0 {\rm for}\, k'=l=0} }
\left( 1 - \exp(2\pi i ( k' \tilde\sigma + l\tilde\rho + j \tilde v) 
\right)^ { \sum_{s=0}^1 \exp{(2\pi i sl)} c^{(r, s)}_2 ( 4k'l -j^2)}, 
\eea
From the threshold integral in \eq{resulint} we can obtain the 
following properties of $\Phit_2'$.
\begin{enumerate}
\item
From \eq{defirsl} it is easy to see that those $SO(2, 3;\ZZZ) = 
Sp(2, \ZZZ)$ transformation
which, acting on the vector $(m_1, n_2, n_1, n_2, b)$  with
$m_1 m_2, n_2, b $ integers and $n_1\in \ZZZ/2$ preserves $m_1$ modulo $2$,
$n_1,, m_2, n_2$ modulo $1$  and $b$ modulo $2$, will be symmetries of 
${\cal I}$ in \eq{rwrint}. 
This defines the subgroup $\tilde G$ \cite{David:2006ji},
thus from \eq{resulint} we see that 
$\Phit_2'$ is a modular form of weight $2$ under the subgroup
$\tilde G$ of $Sp(2, \ZZZ)$. Therefore
\be{modtran2'}
\Phit_2'( (A\Omega + B) (C\Omega + D)^{-1} ) = \det( C\Omega + D)^2
\tilde\Phi_2'(\Omega), \quad\left(
\begin{array}{cc}
A & B\\ C& D \end{array} \right) \in \tilde G.
\ee
Since the $\tilde G$ contains $\tilde H$ given in \eq{subgroup} using the
same arguments as in the case of $\Phit_6$ we see that
$\Phit_2'$ is also invariant under the transformation 
\eq{sghtrans}. 
\item
$\Phit_2'$ has second order zeros at  
\bea{loczero2'}
& & \left( n_2(\tilde \sigma\tilde\rho -\tilde v^2) + b\tilde v + n_1 
\tilde\sigma -
  \tilde\rho m_1 + m_2\right) =0, \\ \nonumber
& & {\rm for}\; m_1\in 2\ZZZ, m_2, n_2\in\ZZZ, n_1\in \ZZZ b\in 2\ZZZ+1, 
m_1n_1 +m_2n_2 + \frac{b^2}{4} = \frac{1}{4}. 
\eea
It also has a second order pole at 
\bea{locpol2'}
&&\left( n_2(\tilde \sigma\tilde\rho -\tilde v^2) + 
b\tilde v + n_1 \tilde\sigma -
\tilde\rho m_1 + m_2\right) =0, \\ \nonumber
& &{\rm for}\, m_1 \in 2\ZZZ, m_2, n_2 \in \ZZZ, n_1\in \ZZZ +\frac{1}{2},
 b\in 2\ZZZ+1, 
m_1n_1 +m_2n_2 + \frac{b^2}{4} = \frac{1}{4}.
\eea
\item
In the limit $\tilde v \rightarrow 0$, $\Phit_2'$ factorizes as
\bea{facprop2'}
\Phit_2' (\tilde\rho, \tilde\sigma, \tilde v)_{\tilde v\rightarrow 0} =
-\frac{\pi^2}{64} \tilde v^2 \frac{\eta(\tilde\rho)^{16}}{\eta(2\tilde\rho)^8}
\frac{\eta(\tilde\sigma)^{16}}{\eta(\tilde\sigma/2)^8}.
\eea
\item
Under the $Sp(2,\ZZZ)$ transformation given in \eq{chgvar}
from the relation of $\Phit_2$ to 
$\Phit_2'$ in \eq{re22'}  it can be shown that  $\tilde\Phi_2'$ is 
related to $\hat\Phi_2$ $Sp(2, \ZZZ)$ modular form of weight 2 by 
\bea{dualprop2'}
\Phit_2'(\tilde\rho, \tilde\sigma, \tilde v) &=&
\tilde\sigma^{-2} 
\hat\Phi_2'\left(\tilde\rho - \frac{\tilde v^2}{\tilde\sigma}, 
\frac{\tilde\rho\tilde\sigma - (\tilde v -1)^2}{\tilde\sigma}, 
\frac{\tilde\rho\tilde\sigma - \tilde v^2 + \tilde v}{\tilde \sigma}
\right), \\ \nonumber 
& =& \tilde \sigma^{-2}
\hat\Phi_2'(  \rho,\sigma, v). 
\eea
where $\hat\Phi_2$ is defined by \cite{David:2006ud}. 
\bea{defphio2'}
\hat\Phi_2(\rho, \sigma, v) &=& - \exp(2\pi i (\rho+ \sigma + v)) \\ \nonumber
&\,& \prod_{b=0}^1
\prod_{r,s=0}^1 \prod_{\stackrel{(k, l)\in \zzz, j\in 2\zzz+ b}
{k, l\geq 0, j<0\, 
\hbox{for}\, k=l=0}} 
\left\{ 1 - (-1)^r \exp( 2\pi i (k \sigma + l\rho + j v)) \right\}
^{ \hat c^{(r,s)}_{2;b} (4kl -j^2)},
\eea
$(\tilde\rho,\tilde\sigma,\tilde v)$ is related $(\rho, \sigma, v)$ by
\eq{chgvar}.
$\hat c^{(r,s)}_{2;b}$ is related to $c^{(r,s)}_2$ by
\be{relhatcc}
\hat c_{2;b}^{(r,s)}( u) = \frac{1}{2} \sum_{r'=0}^1\sum_{s'=0}^1
e^{2\pi i (sr'-rs')/2)} c_{2;b}^{(r',s')} (u). 
\ee
$\hat\Phi_2$ has the factorization property in the limit $v\rightarrow 0$
\be{fprop2'}
\hat\Phi_2(\rho, \sigma, v )_{v\rightarrow 0} = 4\pi^2 v^2 
\frac{\eta(\rho)^{16}}{\eta(2\rho)^8} 
\frac{\eta(\sigma)^{16}}{\eta(2\sigma)^8}.
\ee
\item
Using \eq{dualprop2} and the factorization property 
\eq{fprop2} we see that when $\tilde\rho\sigma -\tilde v^2 + \tilde v \sim 0$
$\tilde \Phi_2'$ factorizes as 
\be{fprops2'}
\tilde\Phi_2'(\tilde\rho, \tilde\sigma, \tilde v) \sim
4\pi^2 ( 2 v -\rho -\sigma)^2 v^2
\frac{\eta(\rho)^{16}}{\eta(2\rho)^8} 
\frac{\eta(\sigma)^{16}}{\eta(2\sigma)^8}.
\ee
\end{enumerate}

Now that we have the properties of the basic modular 
forms $\Phit_2, \Phit_2', \Phit_6$
we can derive the properties of the modular forms $\Phit_0'$ and $\Phit_4'$
which are used in the writing down the dyon partition functions in the
STU model and the FHSV model.

\vspace{.5cm}
\noindent
{\underline{$\Phit_0(\tilde\rho,\tilde\sigma, \tilde v)$}}
\vspace{.5cm}

\noindent
We define $\Phit_0$ as
\be{defaphit0}
\Phit_0  = \Phit_2 \sqrt{ \frac{\Phit_2'}{\Phit_6}}.
\ee

\begin{enumerate}
\item
It is clear that from the above definition this is a modular form of weight 
$0$ under the subgroup $\tilde G$. Since 
$\tilde G$ contains $\tilde H$  it is invariant under the 
transformations \eq{sghtrans}. 
\item
From \eq{loczero6}, \eq{loczero2} and \eq{loczero2'}  and from the definition
\eq{defaphit0} we see that $\Phit_0$ has second order zeros at 
\bea{loczero0a}
& & \left( n_2(\tilde \sigma\tilde\rho -\tilde v^2) + b\tilde v + n_1 
\tilde\sigma -
\tilde\rho m_1 + m_2\right) =0, \\ \nonumber
& & {\rm for}\; m_1\in 2\ZZZ, m_2, n_2\in\ZZZ, n_1\in \ZZZ b\in 2\ZZZ+1, 
m_1n_1 +m_2n_2 + \frac{b^2}{4} = \frac{1}{4}. 
\eea
Further more from the form the locations of the poles of $\Phit_2$ 
given in \eq{locpol2}  and 
$\Phit_2'$  in \eq{locpol2'} and from \eq{defaphit0} we see that $\Phit_0$ has 
second order poles at
\bea{locpol0a}
&&\left( n_2(\tilde \sigma\tilde\rho -\tilde v^2) + 
b\tilde v + n_1 \tilde\sigma -
\tilde\rho m_1 + m_2\right) =0, \\ \nonumber
& &{\rm for}\, m_1 \in 2\ZZZ+1, m_2, n_2 \in \ZZZ, n_1\in \ZZZ, b\in 2\ZZZ+1, 
m_1n_1 +m_2n_2 + \frac{b^2}{4} = \frac{1}{4}.
\eea
It has simple poles at
\bea{locpol0a1}
&&\left( n_2(\tilde \sigma\tilde\rho -\tilde v^2) + 
b\tilde v + n_1 \tilde\sigma -
\tilde\rho m_1 + m_2\right) =0, \\ \nonumber
& &{\rm for}\, m_1 \in 2\ZZZ, m_2, n_2 \in \ZZZ, n_1\in \ZZZ +\frac{1}{2},
 b\in 2\ZZZ+1, 
m_1n_1 +m_2n_2 + \frac{b^2}{4} = \frac{1}{4}.
\eea
\item
From the factorization properties \eq{facprop6} and \eq{facprop2} and \eq{facprop2'}
and from the definition \eq{defaphit0} we see that in the limit
$\tilde v\rightarrow 0$ $\Phit_0$ factorizes as
\bea{facprop0a}
\Phit_0 (\tilde\rho, \tilde\sigma, \tilde v)_{\tilde v\rightarrow 0} =
-\frac{\pi^2}{256} \tilde v^2 \frac{\eta(2\tilde\rho)^{8}}{\eta(\tilde\rho)^4}
\frac{\eta(\tilde\sigma/2)^{8}}{\eta(\tilde\sigma)^4}. 
\eea
\item
\eq{evintp} and \eq{evintp2} ensure that in the 
the product formula for $\Phit_6$, $\Phi_2$ and $\Phi_2'$
given in \eq{defphi6}, \eq{defphi2} and \eq{defphi2'}  the exponents
$c_{6}^{(r,0)}(u) \pm c_6^{(r,1)}$  and $c_{2}^{(r,0)}(u) \pm c_2^{(r,1)}(u)$
are all even integers. The square roots involved in obtaining $\Phit_0$
just make these combinations integers. Therefore we have
\be{intexpphi0}
\frac{K}{\Phit_0 (\tilde\rho,\tilde\sigma, \tilde v) }
=\sum _{\stackrel{m, n, p}{m\geq -1/2, n\geq 1/2} }
e^{2\pi i ( m \tilde\rho + n\tilde\sigma + p \tilde v)} g(m, n, p),
\ee
where the Fourier coefficients $g(m, n, p)$ are integers with $K = - 2^{-10}$. 
$m \in \ZZZ /2 $,  $n \in \ZZZ /2$ and $j \in \ZZZ$. The lower bound in the 
sum in \eq{intexpphi0} and the domains of $m, n, p$ are obtained by 
examining the product representation \eq{defphi6}, \eq{defphi2} and \eq{defphi2'}.
\item
Using \eq{dualprop6}, \eq{dualprop2} and \eq{dualprop2'} it is easily seen that
under the $Sp(2,\ZZZ)$ transformation given in \eq{chgvar} $\Phit_0$ is 
related to $\Phi_0$ the  $Sp(2, \ZZZ)$ modular form of weight 2 by 
\bea{dualprop0}
\Phit_0(\tilde\rho, \tilde\sigma, \tilde v) &=&
 \Phi_0\left(\tilde\rho - \frac{\tilde v^2}{\tilde\sigma}, 
\frac{\tilde\rho\tilde\sigma - (\tilde v -1)^2}{\tilde\sigma}, 
\frac{\tilde\rho\tilde\sigma - \tilde v^2 + \tilde v}{\tilde \sigma}
\right), \\ \nonumber 
& =&
\Phi_0(  \rho,\sigma, v). 
\eea
where $\Phi_0$ is defined by
\be{defphi0a}
\Phi_0(\rho, \sigma, v) = \Phi_2 (\rho,\sigma, v) 
\sqrt{\frac{\hat\Phi_2(\rho,\sigma, v)}{
\Phi_6(\rho,\sigma, v)}}. 
\ee
$(\tilde\rho,\tilde\sigma,\tilde v)$ is related $(\rho, \sigma, v)$ by
\eq{chgvar}.
From \eq{fprop6} and \eq{fprop2} and \eq{fprop2'} it is seen that 
$\Phi_0$ has the factorization property in the limit $v\rightarrow 0$
\bea{fprop0}
\Phi_0(\rho, \sigma, v )_{v\rightarrow 0} & =& 4\pi^2 v^2 
\frac{\eta(2\rho)^8}{\eta(\rho)^4} \frac{\eta(2\sigma)^8}{\eta(\sigma)^4},
 \\ \nonumber
&=& 4\pi v^2 \vartheta_2(\rho)^4 \vartheta_2(\sigma)^4.
\eea
In the last line of the above equation we have written the 
Dedekind-$\eta$ functions in terms of
the Jacobi-$\vartheta$ functions. 
\item
Using \eq{dualprop0} and the factorization property 
\eq{fprop0} we see that when $\tilde\rho\sigma -\tilde v^2 + \tilde v \sim 0$
$\tilde \Phi_0$ factorizes as 
\be{fprops0}
\tilde\Phi_0(\tilde\rho, \tilde\sigma, \tilde v) \sim
4\pi^2 ( 2 v -\rho -\sigma)^2 v^2
 \vartheta_2(\rho)^4 \vartheta_2(\sigma)^4.
\ee
\end{enumerate}

\vspace{.5cm}\noindent
{\underline{$\Phit_4(\tilde\rho, \tilde\sigma, \tilde v)$}}
\vspace{.5cm}

We will not go into the details of the derivation of the properties of $\Phit_4$, but it is
now clear that from the properties of the form $\Phit_6$ and $\Phit_2$ and 
using the definition
\be{defphit4a}
\Phit_4(\tilde\rho,\tilde\sigma, \tilde v) = \sqrt{\Phi_6(\tilde\rho, \tilde\sigma, \tilde v)
\Phi_2(\tilde\rho, \tilde\sigma, \tilde v)}, 
\ee
one can show the properties listed in section 5. of the paper.

\bibliography{n=2}
\bibliographystyle{JHEP}

\end{document}